\documentclass{aa}  
\pdfoutput=1 

\usepackage{graphicx}
\usepackage{tikz}
 
\usepackage{txfonts}
\usepackage[colorlinks=true,
            linkcolor=blue,
            urlcolor=blue,
            citecolor=blue]{hyperref}
\usepackage{bm}
\usepackage{booktabs}
\usepackage{amsmath}
\usepackage{microtype}
\usepackage{hyperref}
\usepackage{wrapfig}
\usepackage{framed}
\usepackage{siunitx}
\usepackage{upgreek}
\usepackage{multirow}
\usepackage{multicol}
\usepackage{textcomp}
\usepackage{breqn}
\usepackage{subcaption}
\usepackage{url}

\makeatletter
\renewcommand*\aa@pageof{, page \thepage{} of \pageref*{LastPage}}

\def\ptrad{{\tt petitRADTRANS}\xspace}
\def\klip{{\tt KLIP}\xspace}

\def\f1{f_{\rm I}}

\def\teff{$T_{\rm eff}$}

\def\h2o{H$_{2}$O\xspace}
\def\ch4{CH$_{4}$\xspace}
\def\nh3{NH$_{3}$\xspace}
\def\co2{CO$_{2}$\xspace}
\def\ph3{PH$_{3}$\xspace}

\def\beq{\begin{equation}}
\def\eeq{\end{equation}}

\def\t2{\tau_{\rm II}}

\def\sigmas0{\Sigma_{\rm s,0}}

\makeatletter
\let\jnl@style=\rm
\def\ref@jnl#1{{\jnl@style#1}}

\def\aj{\ref@jnl{AJ}}                   
\def\actaa{\ref@jnl{Acta Astron.}}      
\def\araa{\ref@jnl{ARA\&A}}             
\def\apj{\ref@jnl{ApJ}}                 
\def\apjl{\ref@jnl{ApJ}}                
\def\apjs{\ref@jnl{ApJS}}               
\def\ao{\ref@jnl{Appl.~Opt.}}           
\def\apss{\ref@jnl{Ap\&SS}}             
\def\aap{\ref@jnl{A\&A}}                
\def\aapr{\ref@jnl{A\&A~Rev.}}          
\def\aaps{\ref@jnl{A\&AS}}              
\def\azh{\ref@jnl{AZh}}                 
\def\baas{\ref@jnl{BAAS}}               
\def\bac{\ref@jnl{Bull. astr. Inst. Czechosl.}}
\def\caa{\ref@jnl{Chinese Astron. Astrophys.}}
\def\cjaa{\ref@jnl{Chinese J. Astron. Astrophys.}}
\def\icarus{\ref@jnl{Icarus}}           
\def\jcap{\ref@jnl{J. Cosmology Astropart. Phys.}}
\def\jrasc{\ref@jnl{JRASC}}             
\def\memras{\ref@jnl{MmRAS}}            
\def\mnras{\ref@jnl{MNRAS}}             
\def\na{\ref@jnl{New A}}                
\def\nar{\ref@jnl{New A Rev.}}          
\def\pra{\ref@jnl{Phys.~Rev.~A}}        
\def\prb{\ref@jnl{Phys.~Rev.~B}}        
\def\prc{\ref@jnl{Phys.~Rev.~C}}        
\def\prd{\ref@jnl{Phys.~Rev.~D}}        
\def\pre{\ref@jnl{Phys.~Rev.~E}}        
\def\prl{\ref@jnl{Phys.~Rev.~Lett.}}    
\def\pasa{\ref@jnl{PASA}}               
\def\pasp{\ref@jnl{PASP}}               
\def\pasj{\ref@jnl{PASJ}}               
\def\rmxaa{\ref@jnl{Rev. Mexicana Astron. Astrofis.}}%
\def\qjras{\ref@jnl{QJRAS}}             
\def\skytel{\ref@jnl{S\&T}}             
\def\solphys{\ref@jnl{Sol.~Phys.}}      
\def\sovast{\ref@jnl{Soviet~Ast.}}      
\def\ssr{\ref@jnl{Space~Sci.~Rev.}}     
\def\zap{\ref@jnl{ZAp}}                 
\def\nat{\ref@jnl{Nature}}              
\def\iaucirc{\ref@jnl{IAU~Circ.}}       
\def\aplett{\ref@jnl{Astrophys.~Lett.}} 
\def\apspr{\ref@jnl{Astrophys.~Space~Phys.~Res.}}
\def\bain{\ref@jnl{Bull.~Astron.~Inst.~Netherlands}} 
\def\fcp{\ref@jnl{Fund.~Cosmic~Phys.}}  
\def\gca{\ref@jnl{Geochim.~Cosmochim.~Acta}}   
\def\grl{\ref@jnl{Geophys.~Res.~Lett.}} 
\def\jcp{\ref@jnl{J.~Chem.~Phys.}}      
\def\jgr{\ref@jnl{J.~Geophys.~Res.}}    
\def\jqsrt{\ref@jnl{J.~Quant.~Spec.~Radiat.~Transf.}}
\def\memsai{\ref@jnl{Mem.~Soc.~Astron.~Italiana}}
\def\nphysa{\ref@jnl{Nucl.~Phys.~A}}   
\def\physrep{\ref@jnl{Phys.~Rep.}}   
\def\physscr{\ref@jnl{Phys.~Scr}}   
\def\planss{\ref@jnl{Planet.~Space~Sci.}}   
\def\procspie{\ref@jnl{Proc.~SPIE}}   

\makeatother

\def\ptp{\ref@jnl{Prog.~Th.~Phys.}}   

\def\({\left(}
\def\){\right)}
\def\<{\left<}
\def\>{\right>}

\begin{document} 
   \title{Impacts of high-contrast image processing on atmospheric retrievals.}
    \titlerunning{High-Contrast Imaging Retrievals}
    \authorrunning{Nasedkin et al.}
   \subtitle{}

    \author{E. Nasedkin
          \inst{1}\fnmsep\thanks{Corresponding Author}
          \and P. Mollière\inst{1}
          \and J. Wang\inst{2}
          \and F. Cantalloube\inst{3}
          \and L. Kreidberg\inst{1}
          \and L. Pueyo\inst{4}
          \and T. Stolker\inst{5}
          \and A. Vigan\inst{3}
          }

    \institute{Max-Planck-Institut für Astronomie, Königstuhl 17, 69117 Heidelberg, Germany\\
              \email{nasedkin@mpia.de}
              \and Center for Interdisciplinary Exploration and Research in Astrophysics (CIERA) and Department of Physics and Astronomy, Northwestern University, Evanston, IL, 60208, USA, 
              \and Aix Marseille Univ., CNRS, CNES, LAM, Marseille, France
              \and Space Telescope Science Institute (STSci), Baltimore, MD 21218, USA
              \and Leiden Observatory, Leiden University, P.O. Box 9513, 2300, RA Leiden, The Netherlands
              }
   \date{Received 00-00-00; accepted 00-00-00}

 \abstract{
  Many post-processing algorithms have been developed in order to better separate the signal of a companion from the bright light of the host star, but the effect of such algorithms on the shape of exoplanet spectra extracted from integral field spectrograph data is poorly understood. 
  The resulting spectra are affected by noise that is correlated in wavelength space due to both optical and data processing effects. 
  Within the framework of Bayesian atmospheric retrievals, we aim to understand how these correlations and other systematic effects impact the inferred physical parameters.
  We consider three algorithms ({\texttt KLIP}, {\texttt PynPoint} and {\texttt ANDROMEDA}), optimizing the choice of algorithmic parameters using a series of injection tests into archival SPHERE and GPI data of the \object{HR~8799} system.
  The wavelength-dependent covariance matrix is calculated to provide a measure of instrumental and algorithmic systematics.
  We perform atmospheric retrievals using {\texttt petitRADTRANS} on optimally extracted spectra to measure how these data processing systematics influence the retrieved parameter distributions.
  The choice of data processing algorithm and parameters significantly impact the accuracy of retrieval results, with the mean posterior parameter bias ranging from 1 to 3 $\sigma$ from the true input parameters.
  Including the full covariance matrix in the likelihood improves the accuracy of inferred parameters, and cannot be accounted for using ad hoc scaling parameters in the retrieval framework.
  Using the Bayesian information criterion and other statistical measures as a heuristic goodness-of-fit metrics, the retrievals including the full covariance matrix are favoured when compared to using only the diagonal elements.}
   

   \maketitle
%

\section{Introduction}\label{sec:intro}

\begin{table*}[ht]
    \centering
    \caption{Epochs of SPHERE ([1] \cite{zurlo_first_2016} 60.A-9249(C)) and GPI ([2] \cite{greenbaum_gpi_2018}  GS-2015B-Q-500-1394) data used.}
    \begin{tabular}{r|lccccccc}
        \toprule
        \textbf{Instrument} &\textbf{Date} & \textbf{Band} & $\boldsymbol{\lambda/\Delta\lambda}$ & \textbf{Field Rotation [\textdegree]} & \textbf{Med. Seeing [as]} & \textbf{DIT [s]} & \textbf{NEXP} & \textbf{Ref.}\\
        \midrule
        \multirow{2}{*}{SPHERE} & \multirow{2}{*}{2014-08-12} & \multirow{2}{*}{YJH} & \multirow{2}{*}{29}  & 29.65 & \multirow{2}{*}{0.87} & 100 & 32 & \multirow{2}{*}{[1]}\\
                                &                             &                      &                      & 15.37 &      & 60  & 48 &   \\
        GPI                     & 2016-09-19                  & H                    & 45                   & 20.93 & 0.97 & 60  & 60 & [2]\\
        \bottomrule
    \end{tabular}
    \label{tab:data}
\end{table*}

The field of high contrast imaging (HCI) has advanced dramatically over the last two decades. 
From the first detection of \object{2M~1207~b} \citep{Chauvin20052M1207b} to the ongoing large surveys such as GPIES \citep{nielson2019_gpies} and SHINE \citep{Desidera2021SHINEI,Langlois2021SHINEII, Vigan2021SHINEIII}, we have seen improvements in instrumentation, adaptive optics and data processing that have led to the discovery of numerous new exoplanets.
Such surveys have established the rarity of giant, widely separated companions, finding that $<10$\% of high-mass stars have planetary mass companions between 10-100 AU.
However, much of this work has remained focused on the detection of new companions at higher contrast ratios and smaller angular separations.
The spectroscopic characterization of known planets has seen less dedicated effort - to date there has not been a uniform survey of known objects to present an homogeneous sample of spectroscopic measurements. 
This can lead to systematic discrepancies between measurements made with different instruments, and challenges in fitting datasets with different spectral resolutions \citep{xuan2022_cloudybds}.
Such biases may impact the conclusions made from population studies, such as the exploration of the C/O ratios of the directly imaged planet population in \cite{Hoch2022_CO}.
Individual characterisation efforts have nevertheless lead to intriguing findings: measurements of water and carbon monoxide abundances in the \object{HR~8799} planets \citep{konopacky_2013_coh2o,lavie_helios-retrieval_2017, wang_chemical_2020}, precise constraints on the C/O ratio and metallicity of \object{$\beta$ Pictoris b} \citep{gravity_collaboration_peering_2020}, measurements of isotope ratios \cite{ZhangY2021_isotopes} and the detection of a dusty envelope around \object{PDS~70~b} and c \citep{Wang2021PDS70, Benisty2021_pds70c}. 

This characterisation work remains challenging.
Extensive post-processing is required to extract the faint signal of the target. 
Even in the most careful analysis there usually remain systematic biases from both instrumental and processing effects.
Integral field spectrograph (IFS) measurements introduce correlated noise as a function of wavelength due to pixel cross talk, interpolation effects and imperfect adaptive optics correction (speckles).
\cite{greco_brandt_2016} provide a method for empirically estimating the correlation from IFS data. 
They demonstrated that accounting for such correlations is necessary when analysing exoplanet atmospheres, and failing to do so will lead to biased and overconfident posterior distributions on measured parameters.
Efforts such as the Exoplanet Imaging Data Challenge \citep{CantalloubeEIDC} explored the detection abilities of a suite of HCI algorithms but a systematic algorithmic comparison for spectral characterisation has not yet been performed.

Most post-processing techniques are based on Angular Differential Imaging (ADI)  \citep{Marois_adi_2006,marois_direct_2008}, where the telescope is pupil stabilized and the field is allowed to rotate. 
This provides differential motion of the planet over the course of the observations and allows for the removal of stellar speckles by derotating and stacking the resulting images. 
Ongoing development of this method has been largely driven by the goals of increasing sensitivity at small angular separations.
To this end, different algorithms have been developed to maximize the information available in imaging datasets, leveraging spatial and spectral information in order to separate the faint planet signal from the bright host star.
\cite{kiefer2021} explored how different approaches impact the $S/N$ of IFS observations, but did not examine the impact of the processing on the extracted spectral shape.

The use of atmospheric retrievals to study directly imaged planets is relatively new, with only a small but growing selection of targets being subject to such an analysis \citep[e.g.,][]{Lee_2013_HR8799bret, lavie_helios-retrieval_2017, molliere_retrieving_2020, gravity_collaboration_peering_2020,brown-sevilla2022, Whiteford2023_51Erib}.
While the effects of systematics are well understood for transmission spectroscopy using HST \citep{ih_2021_systematics}, with significant efforts extending this to JWST \citep{Barstow_2015_jwstsystematics,rocchetto_exploring_2016,Lacy_2020_jwstbiases}, the impact of systematic uncertainties in ground-based high-contrast data on atmospheric retrievals has not been thoroughly explored.
Even in the era of \textit{JWST}, understanding systematics is critical to interpreting model fits to data. 
Ground based observations will remain a key component of this understanding due to their higher spectral and angular resolution that cannot yet be achieved from space.

In this work we explore the systematic effects introduced through high-contrast data processing on the retrieval of atmospheric parameters.
The details of our example datasets used are described in Section \ref{sec:obs}.
Section \ref{sec:data} outlines our methods, exploring the different algorithms we test in Section \ref{sec:algs}, together with the measurement and interpretation of the covariance matrix in Sections \ref{sec:covar} and \ref{sec:covimpact}.
We determine the optimal parameters for spectral extraction through the injection of synthetic companions into the data in Section \ref{sec:Injections}.
The results of our retrieval comparisons are described in Section \ref{sec:retrievals}, while the implications and limitations of these results are discussed in Section \ref{sec:discussion}.

\section{Observations}\label{sec:obs}
While the first goal of our study is to demonstrate the effects that post-processing algorithms can have on inferred atmospheric parameters for general high-contrast spectroscopy, we still had to select demonstration datasets. We chose GPI and SPHERE observations of the well-known four-planet system in HR~8799 \citep{marois_direct_2008,marois_images_2010}, where discrepancies between GPI and SPHERE datasets, covering the same wavelength range, had already been noted \citep{lavie_helios-retrieval_2017, molliere_retrieving_2020}.
HR~8799 has seen extensive photometric and spectroscopic observing campaigns, e.g. \citep{konopacky_2013_coh2o, zurlo_first_2016, lavie_helios-retrieval_2017, greenbaum_gpi_2018, gravity_collaboration_first_2019, molliere_retrieving_2020, wang_chemical_2020, Ruffio2021_HR8799Osiris, WangJi2023_HR8799c}. 
The importance of this system, together with the abundance of high contrast data from multiple instruments make it an ideal object of study for our purposes. 
As a benchmark target, the companions have luminosity and spectra typical of this class of low surface gravity object and are representative of the current directly imaged exoplanet population. 

\subsubsection*{SPHERE}
The SPHERE data were taken during the commissioning run of the SPHERE instrument \citep{beuzit_sphere_2008, beuzit_sphere_2019} in 2014, and were originally presented in \cite{zurlo_first_2016}.
It remains the best YJH band spectrum of HR~8799 to date in terms of signal-to-noise and spectral resolution.
IFS frames in the YJH band were taken with a series of both 60 s and 100 s integrations, using pupil-stabilized observations to allow for ADI post-processing.
Total field rotations of 15.37\textdegree\ and 29.65\textdegree\ were observed for the 60 s data cube and for the 100 s data cube, respectively.
To compensate for the difference in exposure time, we multiply each 60 s exposure by a factor of 100/60, in order to process the data as a whole.
We rereduced the SPHERE data using the pipeline described in \cite{vigan-sphere-pipeline-2020}: details of which are described in Appendix \ref{sec:sphereappendix}.

\subsubsection*{GPI}
The GPI \citep{macintosh_first_2014} observations of HR8799 were originally published in \cite{greenbaum_gpi_2018} and were taken on 2016-09-19, 2013-11-17 and 2013-11-18 for the H, K1 and K2 bands respectively.
As with the SPHERE data, the telescope was pupil-stabilized to take advantage of ADI post-processing.
These were reduced using the standard GPI reduction pipeline (version 1.4.0). The median seeing of the observations was 0".97; the observing conditions are more thoroughly described in \cite{ingraham_gemini_2014}.
While data were taken in the H, K1 and K2 bands of GPI, we only consider the H band observations due to the low $S/N$ of the K-band observations.
The observations from both GPI and SPHERE are summarized in Table \ref{tab:data}.

\subsection{Data preprocessing}
In order to reduce the systematic variation between the datasets, we first rereduced the data with up-to-date pipelines. 
For both the SPHERE and GPI datasets, we then preprocess the IFS cubes using the Vortex Image Processing (VIP) library in order to select the optimal frames for further ADI processing.
The \verb|cube_detect_badfr_correlation| function computes the similarity between each frame and a reference frame in order to identify frames that are outliers when compared to the rest of the sequence.
We choose the frame which maximizes the mean similarity of all frames as the reference frame, and remove the most different 12\% of frames from each the SPHERE and GPI datasets.
Such variation in the data is typically due to changing observing conditions, introducing effects into the data such as the low-wind effect \cite{Milli2018_lowwind} or the wind-driven halo \cite{Cantalloube2020_winddrivenhalo}
This threshold is sufficient to remove frames which are significantly outlying and visually show differences when compared to the a typical frame.
This leaves 69 ADI frames for the SPHERE data set, and 51 for the GPI H-band dataset.

\begin{table}[t!]
    \centering
    \caption{Stellar properties of HR~8799 A.}
    \begin{tabular}{lll}
        \toprule
        \multicolumn{3}{c}{\textbf{HR~8799}}\\
        \midrule
        \textbf{Parameter} & \textbf{Value} & \textbf{Note}\\
        \midrule
        $\alpha$ [J2000]        & $23^{h}\:07^{m}\:28.7157^{s}\pm0.0685^{s}$ & [1] \\
        $\delta$ [J2000]        & $+21$\textdegree$\:08^{'}\:03.3021^{''}\pm0.0799^{''}$& [1] \\
        $\upmu_{\alpha}$ [mas/yr] & $108.301\pm0.168$ & [1] \\
        $\upmu_{\delta}$ [mas/yr] & $-49.480\pm0.152$ & [1] \\
        $\bar{\omega}$ [mas]    & $24.2175\pm0.0881$  & [1] \\
        $d$ [pc]                & $41.2925\pm0.1502$ & [1]\\
        RV [km s$^{-1}$]        & $-12.60\pm1.4$ & [2] \\

        Spectral Type           & F0+VkA5mA5 C & [3] \\
        \midrule
        \teff [K]               & $7200\pm50$ & [4] \\
        log $g$ [cgs]           & $3.0\pm0.25$ & [4] \\
        \textrm{[Fe/H]} [dex]   & $0.0\pm0.2$ & [4] \\
        $R_{\star}$ [R$_{\sun}$]& $1.496\pm0.0054$  & [4] \\
        $L_{\star}$ [L$_{\sun}$]& $5.230\pm0.0498$ & [4] \\
        C/O                     & $0.54_{-0.09}^{+0.12}$ & [5] \\
        \bottomrule
    \end{tabular}\\
    \begin{flushleft} 
    \textbf{Notes.} [1] \cite{gaia_collaboration_gaia_2018}. [2] \cite{gontcharov_hipparcos_rv}. [3] \cite{gray_hr8799_stellar_type}. [4]  {\tt BT-NextGen} best-fit to photometry \citep{hasuchildt_1999_btnextgen}. [5] \cite{wang_chemical_2020}. 
    
    HR~8799 is a $\lambda$ Bo\"{o}tis star, for further discussion see \cite{Molliere2022_formation}.
    \end{flushleft}
    \label{tab:stellar_data}
\end{table}

\subsection{Stellar model for flux extraction}\label{sec:hr8799star}
In order to obtain the absolute flux of the companions we use a model of the stellar spectrum to flux-calibrate the contrast measurements.
HR8799 is an F0+VkA5mA5 C star \citep{gray_hr8799_stellar_type} located $41.3\pm0.2$ pc \citep{gaia_collaboration_gaia_2018}.
Stellar photometry of \object{HR~8799} from WISE and 2MASS is used to fit model stellar spectrum \citep{Cutri2013_WISE,Cutri2003_2Mass}.
We exclude data points beyond 5 $\upmu$m  so that the fit is not impacted by the infrared excess from the debris disk \citep{Su_2009_hr8799debrisdisk, Faramaz_2021_hr8799debrisdisk}.
Using the {\tt species} package \citep{stolker_species}, we fit a {\tt BT-Nextgen} model to the photometry within our wavelength range of interest.
The best-fit model has parameters of \teff = 7200 K, log g = 3.0 and [Fe/H] = 0.0, slightly cooler than the models used in previous studies \citep{zurlo_first_2016,greenbaum_gpi_2018}. 
The full set of stellar parameters is listed in table \ref{tab:stellar_data}.
This spectrum is normalized to a 10 pc distance.
The model is convolved to the instrumental spectral resolutions and binned to the instrumental wavelength channels to allow for spectrophotometric calibration of contrast measurements.

In order place measurements of planet properties in context it is also necessary to understand the properties of the host star. 
\cite{wang_chemical_2020} used HARPS observations to directly measure the C and O abundances of the star, finding a C/O ratio of $0.54^{+0.12}_{-0.09}$.
HR~8799 is a $\lambda$ Bo\"{o}tis star, known to be depleted in iron \citep{Gray2002_hr8799lamboot}.
Consistent with this, the authors fit \ion{Fe}{I} and \ion{Fe}{II} lines, finding a metallicity of $\mathrm{[Fe/H]} = -0.52\pm0.08$. 
Both of the carbon and oxygen abundances were measured to be consistent with solar composition, suggesting that the iron metallicity is not representative of the bulk stellar composition, and that our {\tt BT-NextGen} is still applicable. 
At the low spectral resolution considered in this study the metallicity does not significantly impact the SED of the star and variations in its measurement will not affect the calculation of the results, though will ultimately impact the  the context -- and thus interpretation -- of planetary metallicity measurements.


\begin{figure*}
    \centering
    \includegraphics[width=\linewidth]{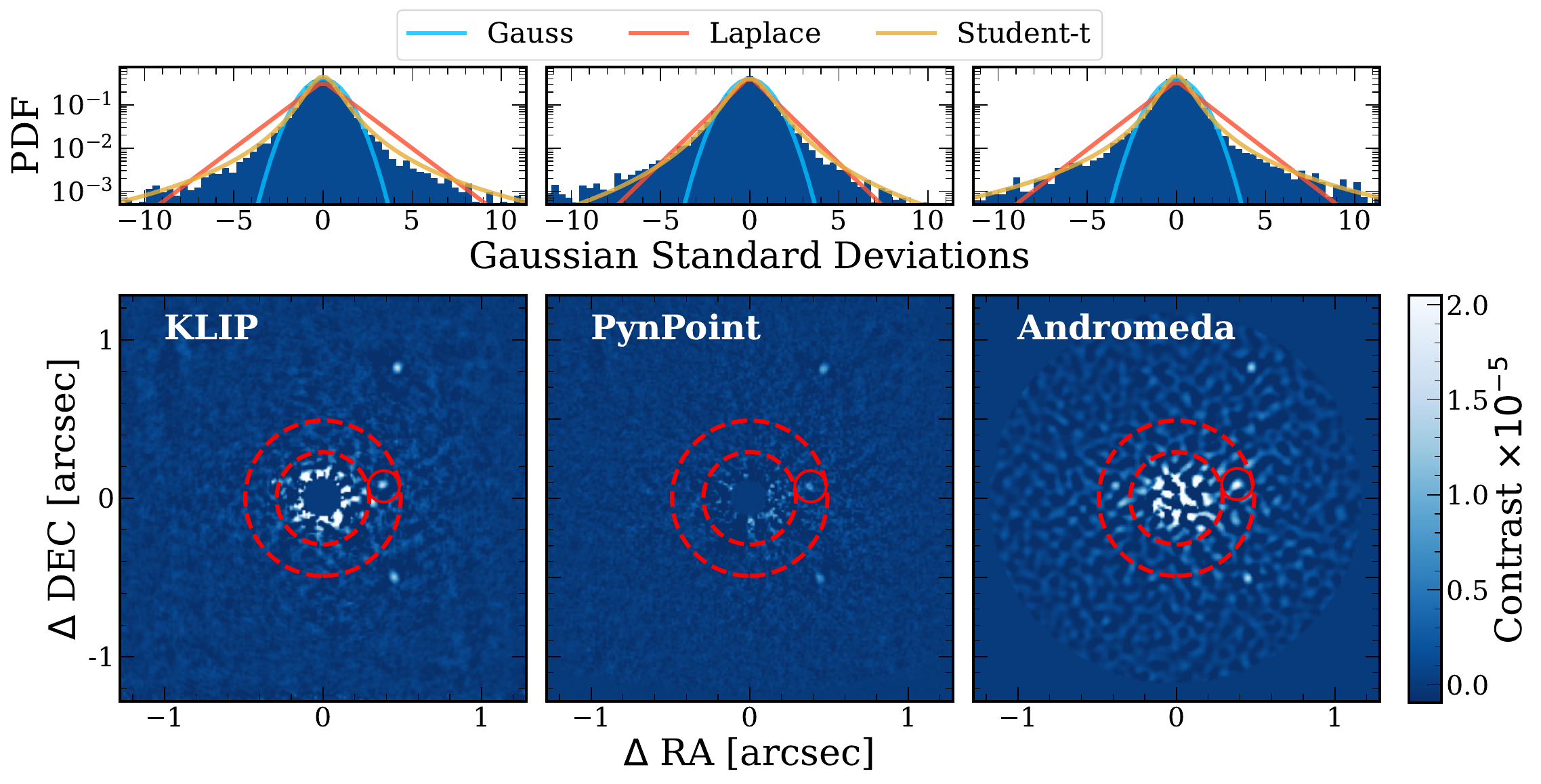}
    \caption{A single wavelength channel from the GPI H-band data of the HR~8799 system, post-processed with \texttt{KLIP} (left), \texttt{PynPoint} (center) and \texttt{Andromeda} (right). 
    The solid red circle denotes the position of HR~8799~e as computed using the \texttt{pyKLIP} astrometry module. Marked in red dashed lines are the regions which are compared when finding the mean correlation between different wavelength channels. 
    At the top of the figure, histograms of the residuals are plotted in units of $\sigma$. 
    The light blue line is a Gaussian fit with the width defined as the Gaussian standard deviation of the residual frame below. }
    \label{fig:ifuframes}
\end{figure*}

\section{ADI Data Processing}\label{sec:data}
The atmospheric properties of directly imaged exoplanets are presently  accessible only through their thermal emission.
For directly imaged planets, these spectra are usually obtained through low to moderate resolution IFS instruments equipped with coronagraphic optics.
IFS data is complex, with a large array of systematic and random noise effects imprinted onto the data.
Cross talk between neighbouring pixels due to optical effects \citep{Antichi2009_BIGREIFUCrossTalk, Larkin2014_GPI} and scattered light can introduce correlations in wavelength space. 
Once the data has been reduced from raw detector frames to data cubes, quasi-static stellar speckles -- light of the host star scattered by the telescope optics -- is the dominant noise source \citep{Marois2005_speckles,Marois2008_NoiseLimits}.
ADI processing is used to remove the stellar PSF and speckle noise, taking advantage of the stability of the PSF over time \citep{Marois_adi_2006}. 
ADI exploits the rotation of the planet through the frame, which produces a signal that is different from the stellar speckles, which remain fixed in position. 
By derotating and stacking the images, the residual speckles following post-processing are averaged out, while the planet signal is enhanced.
The stability assumption is not without flaws, as the PSF varies due to thermal variation in the telescope, short and long-term atmospheric changes and more \citep{Milli2016_specklelifetime}, but in practice it is robust enough to allow for planet detection.
Obtaining an exoplanet spectrum is generally achieved by applying an ADI algorithm to each spectral channel on a 4D cube of IFS data. Modern ADI processing is more sophisticated than simply derotating and stacking the images, but the algorithms generally fall into three broad categories:

\textit{Speckle subtraction} methods attempt to directly subtract the residual stellar speckles from each frame of the image cube.
The planet signal is then measured either through aperture photometry or through fitting a model of the PSF to the signal and minimizing the residuals.
This is the most commonly used method, and includes algorithms such as (Template) Locally Optimized Combination of Images (\verb|LOCI| and \verb|TLOCI|, \citealt{lafreniere_2007_loci,Maire_2012_loci,Marois_2014_TLOCI}), low rank plus sparse decomposition (\verb|LLSG|, \citealt{gonzalez_2016_llsg}) and various implementations of principal component analysis (PCA) based methods, including  Karhunen-Lo\`{e}ve Image Projection (\verb|KLIP|, \citealt{Soummer_2012_KLIP}) and Standardized Trajectory Intensity Mean (\verb|STIM|, \citealt{parait_2019_stim}). 
Such PCA based methods construct an ordered library of principal components of the data: low orders describe the most important components of the stellar PSF, while higher orders describe high frequency noise. 
By building a library to describe the host star PSF, it can be more effectively subtracted from each frame before stacking the images, improving the $S/N$ of the companion.

\textit{Inverse methods} such as \verb|ANDROMEDA| \citep{Mugnier_2009_AND, Cantalloube2015}, \verb|PACO| \citep{Flasseur_2018_PACO} and \verb|TRAP| \citep{Samland_2021_trap} use likelihood minimization to directly estimate the position and contrast of proposed signal at each point in the field. 
To do this, a parameterized forward model of the companion signature is fit to the data, and the parameters are optimized through a likelihood minimization process.
This yields a statistical interpretation of the residuals, and provides confidence region estimates that provide a metric for detection significance, under varying assumptions of the noise properties of the data.

Finally, \textit{supervised machine learning} methods \citep{gonzalez_2018_supervised,yip_2019_deepl,Gebhard_2022_halfsibling} are trained on large sets of data with injected targets and learn how to identify the presence of a companion in an image. 
These methods typically only produce a binary maps where a planet is either detected or not, and do not measure the strength of the planet signal.
Thus these methods have not generally been used for exoplanet characterization.

\subsection{Post-processing algorithms}\label{sec:algs}
We chose to compare three widely used ADI techniques in order to determine the impact of such post-processing on the spectral shape and noise properties of the extracted exoplanet spectrum. 
In order to compare a diverse range of techniques we chose to use \verb|KLIP| and \verb|PynPoint|, which are different flavours of PCA based speckle subtraction methods, and \verb|Andromeda|, which is an inverse method.
As our goal is to understand the impact of systematic effects, we chose these algorithms for their broad community use, typifying the effects likely present in existing work.
A more complete examination of the diversity of algorithms, including spectral differential imaging (SDI) and ADI+SDI algorithms will be explored using a larger set of data in a forthcoming publication based on Phase 2 of the Exoplanet Imaging Data Challenge.

In order to assess our choice of algorithm, we will compare extractions of known injected spectra at different positions and contrasts in order to optimise the parameter selection for extracting the true spectrum.
In this section we will present the specific steps we took to reduce SPHERE and GPI datasets of the HR~8799 system using each of these algorithms. While a wide range of parameters were explored, Table \ref{tab:algoparams} summarizes the parameter choices used in this analysis for each algorithm.

\subsubsection*{KLIP}
\verb|KLIP| is a PCA-based speckle subtraction algorithm, described in \cite{Soummer_2012_KLIP, wang_klip} and \cite{pueyo_klip}.
A Karhunen-Lo\`{e}ve transform of an optimized combination of reference images is used to define the basis of eigenimages, onto which the science frames are projected.
Often this set of reference images is derived from the science observations, but in principle can be any representative measurements of the PSF.
Mathematically, this is equivalent to building the basis of principal components.
This projection is subtracted from the science frames in order to produce the final residual image.
A forward model of the PSF is then injected in order to measure the position and contrast of a detected companion.

Our choice of KLIP parameters is guided by \citet{pueyo_klip} and \citet{greenbaum_gpi_2018}.
For this study, we use {\tt KLIP} in ADI mode. 
Comparison tests showed that the full ADI+SDI mode provided modest increases in $S/N$ at low contrasts, but the overall shape of the spectrum remained similar.
We set a region around the proposed location of the planet extending 13 pixels radially in each direction, and {18\textdegree} on either side of the planet.
The flux overlap parameter is used to set the aggressiveness of the subtraction, using a value of 0.1.
Fixing these parameters may result in sub-optimal spectral extraction, particularly at very small separations where the rotational movement of the planet through the frame is small.
However, we are primarily concerned with the overall trends in the spectral extractions and noise properties across different tools, and do not attempt to fine tune each algorithm for each individual injections.

We use the \verb|pyKLIP| astrometric measurement tools to compute the location of each target within the field of view, which is used to provide our initial estimate for the planet position for each of the algorithms we consider.
The extracted spectrum is highly sensitive to the inferred companion position, and so we use the \verb|KLIP| astrometry as the location for all three algorithms.
\cite{pueyo_klip} outlines the procedure to extract the spectrum from \klip processed data using the forward model extraction tool.
For each target at each wavelength, a forward model is generated from the unsaturated PSF obtained during the observation. KLIP processing is then applied to subtract the stellar PSF and measure the contrast of the companion.
This is converted into a flux measurement using the \verb|BT-NextGen| model of the host star spectrum from Section \ref{sec:hr8799star}. 

\subsubsection*{PynPoint}
\verb|PynPoint| is a Python package designed for high contrast imaging data processing\citep{amara_pynpoint,stolker_pynpoint}. 
The standard PSF subtraction method used in the package is based on full-frame PCA.
We process each wavelength channel of the IFS data independently, filtering for bad pixels and running ADI-PCA on each stack of images.
In contrast to \verb|KLIP|, which builds a model of principal components in a local region near the planet, \verb|PynPoint| builds its PC library from the full available field of view.
The central 0."12 of each frame is masked out, due to the large residuals close to the host star.

Following the PSF subtraction, a PSF model with negative flux is injected at the position of the planet of interest, which is known from previously computed \verb|KLIP| astrometry.
The PSF model for the planet is simply the stellar PSF, which is either derived from satellite spots (for GPI data) or from unocculted observations of the host star (for SPHERE data).
The position and magnitude of the negative planet are iteratively fit to the data using a simplex minimization routine to minimize the $\chi^{2}$ between the PSF model and the data.
The minimisation is considered within an \verb|aperture| with a radius of 4 pixels around the proposed location of the planet.
The iteration continues until a \verb|tolerance| of 0.01 is reached for both the planet position and contrast in magnitude units. 
We allow the planet position to vary by up to 3 pixels (\verb|offset|) from the initial estimate from \verb|pyKLIP| astrometry.
This produces a best-fit value of the position and contrast-magnitude of the planet.
While we allowed the number of principal components used to vary from 1 to 25, we found that the extraction quality degraded substantially after 15 components, which sets the upper bound we present in this work.
This is then converted from magnitude to contrast, and multiplied by the \verb|BT-NextGen| stellar model of Section \ref{sec:hr8799star} to find the absolute flux of the planet.

\begin{table}
    \centering
    \begin{tabular}{l|c}
    \toprule
    \textbf{Parameter} & \textbf{Value}\\
        \midrule
        \multicolumn{2}{c}{ \texttt{pyKLIP} }\\
        \midrule
            \verb|nPC|           & $1-25$\\
            \verb|flux_overlap|  & 0.1   \\
            \verb|highpass|      & True  \\
            \verb|maxnumbasis|   & 150    \\
            \verb|mode|          & ADI   \\
        \midrule
        \multicolumn{2}{c}{ \texttt{PynPoint} }\\
        \midrule
            \verb|nPC|          & $1-15$\\
            \verb|merit|        & Gaussian  \\
            \verb|aperture|     & 4 px\\
            \verb|tolerance|    & 0.01\\
            \verb|cent_size|    & 0."12\\
            \verb|offset|       & 2 px\\
        \midrule
        \multicolumn{2}{c}{ \texttt{ANDROMEDA} }\\
        \midrule
            \verb|filtering_frac|   & 0.35, 0.30 \\
            \verb|min_sep|          & 0.45 $\lambda$/d, 0.25 $\lambda$/d\\
            \verb|width|            & 0.8 $\lambda$/d, 1.2 $\lambda$/d\\
            \verb|iwa|              & 2.0 $\lambda$/d, 1.0 $\lambda$/d \\
            \verb|owa|              & 60/$S$, 45/$S$\\
            \verb|opt_method|       & lsq\\
    \bottomrule
    \end{tabular}
    \caption{Parameters used for each of the algorithms considered. Parameters that were not varied were set based on previously reported values in literature \citep{pueyo_klip,greenbaum_gpi_2018,Cantalloube2015}. For {\texttt ANDROMEDA}, the first column of parameters was used for the SPHERE data, and the second column for the GPI data. The oversampling parameter $S$ is defined in eqn. \ref{eqn:oversamp}. Further information about each of these parameters is available in the documentation of each package.}
    \label{tab:algoparams}
\end{table}
\subsubsection*{Andromeda}
\verb|ANDROMEDA| (ANgular DiffeRential OptiMal Exoplanet Detection Algorithm) is a maximum likelihood estimation algorithm for ADI data, and estimates the position and flux of point sources within the field of view \citep{Mugnier_2009_AND,Cantalloube2015}.
We run the VIP implementation of the algorithm on each wavelength channel independently, and combine the extracted contrast and standard deviation to build the planet spectrum.
\verb|ANDROMEDA| begins by high-pass filtering the data to remove large spatial scale structure from each data frame. 
This step induces signal loss, {and we chose a value of 0.3 for the filtering fraction parameter, leading to a $\sim$20\% energy loss as in Figure 1 of \cite{Cantalloube2015}.
We calculate the \verb|oversampling| parameter for each wavelength channel to ensure the sampling is constant across wavelength, and additionally use this parameter to determine the outer working angle, which is provided in $\lambda/D$. The oversampling parameter, $S$ is defined to be
\begin{equation}\label{eqn:oversamp}
    S = \frac{\lambda}{2\sigma_{\rm px}D}
\end{equation}
for pixel scale $\sigma_{\rm px}$, telescope diameter $D$ and wavelength $\lambda$.

Internally to \verb|ANDROMEDA| pairs of images are chosen such that they are as close together in time as possible to preserve speckle self-similarity, while still ensuring movement of the proposed companion in order to avoid self-subtraction.
This is done on an annular basis, as the motion of the planet depends on the separation from the host star.
A scaling factor $\gamma$ is fit using a least squares method to ensure that the mean of the intensity distribution of both images in the pair is equal.
Using the assumption that the residual noise is white and Gaussian, \verb|ANDROMEDA| then can perform a likelihood test to identify the presence of a companion, by minimizing the difference between the residuals and a model of the companion signal.

Among the outputs of this algorithm are a contrast map, where each pixel represents the contrast of the planet if it was centred on that pixel, and a standard deviation map, specifying the uncertainty associated with each contrast estimate. 
This is different from the output of a speckle-subtraction algorithm, where the flux of an object must be estimated through aperture photometry or via fitting a PSF model to the residuals.
To extract the spectrum, we sum the $S/N$ map along the wavelength axis, and identify the maximum $S/N$ pixel in a 10 pixel box around the known position of the planet. 
We then use this location to measure the contrast and standard deviation as a function of wavelength.

\subsection{Spectral covariance estimation}\label{sec:covar}
Both high contrast imaging and IFS observations present challenges when deriving robust uncertainty estimates, as correlations are naturally present in the data.
Due to aberrations in the telescope optics, imperfect correction for atmospheric turbulence from the adaptive optics systems, and imperfect stellar PSF subtraction, speckles from the stellar PSF are the dominant noise source for AO assisted, high-contrast data sets \citep{Marois_adi_2006}. 
These speckles move radially as a function of wavelength, scaling with the size of the stellar PSF.
This induces a correlation between wavelength channels, as a speckle will take several channels of movement to pass over a pixel at a fixed separation.
Crosstalk -- light from a single lenslet in the lenslet array diffracting into neighbouring channels -- will also couple these channels.
Finally, as noted in \cite{Ruffio2021_HR8799Osiris}, additional correlation can be introduced through the interpolation of the 4D $\left(\lambda,t,x,y\right)$ spectral cube during reconstruction from the detector images.
This interpolation to a fixed wavelength grid guarantees the correlation of the noise in the IFS cubes, as noise in neighbouring detector pixels will be interpolated to build the IFS spaxels.

\cite{greco_brandt_2016} demonstrate the necessity of accounting for these correlated errors when retrieving physical properties from IFS data.
If these correlations are not accounted for, they find that the retrieved confidence intervals are both artificially small and unreliable, often excluding the true parameter values at $>95$\% confidence.
This was reinforced by \cite{ih_2021_systematics}, where they explored the impact of correlated noise on atmospheric retrievals for transiting planets, finding that the assumption of non-correlated noise will lead to biased posteriors and overfitting of the data.

\subsubsection*{Measuring Noise Correlation}
\cite{greco_brandt_2016} introduce a procedure for empirically measuring the correlation in IFS data sets, and demonstrated the importance of including the full covariance matrix when fitting IFS spectra. 
In this work we extend their method by measuring the spectrum of injected planets and the resulting covariance, as opposed to the parameterised noise instance used in their work.
This allows us to explore how the noise properties vary across instruments and over different post-processing methods.
For each PSF-subtracted dataset we compute the average correlation within a 6 pixel wide annulus centered at the separation of the companion of interest. 
As in their work, we find that the correlation matrix does not depend strongly on the width of this annulus.
The companion itself is masked out, leaving only residual noise.
Such an annulus is chosen in order to maintain consistent noise properties in the sample of pixels: in general the noise varies more strongly with radius than with position angle.
Work such as \cite{Gebhard_2022_halfsibling} explores choosing more a more representative sample of pixels to describe the noise at the location of the planet, but such methods are computationally expensive, and we see little azimuthal asymmetry in the residuals shown in Figure \ref{fig:ifuframes}. 

Within the annulus, we compute the elements of the correlation matrix, $\psi_{ij}$ as 
\begin{equation}\label{eqn:cor}
    \psi_{ij} = \frac{\left<I_{i}I_{j}\right>}{\sqrt{\left<I_{i}^{2}\right>\left<I_{j}^{2}\right>}} = \frac{C_{ij}}{\sqrt{C_{ii}C_{jj}}},
\end{equation}
where $\left<I_{i}\right>$ is the mean pixel intensity in the $i^{th}$ spectral channel, and $C_{ij}$ is the covariance between the two channels.

\subsubsection*{Estimating Uncertainties}
In order to compute the covariance matrix from the correlation matrix, we must know the diagonal, or uncorrelated elements of the covariance matrix.
Several methods of measuring the photometric uncertainty were considered.
We estimate the uncorrelated error in each wavelength channel by combining the photometric uncertainty of the stellar PSF, $\sigma_{\rm star}$,with the residual noise at the location of the planet, $\sigma_{\rm residual}$.
We include the stellar uncertainty because near the edges of the bands in which spectra are observed, the filter transmission drops and atmospheric absorption increases, resulting in an increase in the uncertainty on the host star photometry.
To measure the uncertainty on the stellar photometry we measure the standard deviation of the background in an annulus far from the stellar PSF in each wavelength channel, and use this to calculate the signal to noise.
This represents an optimistic estimate of the stellar uncertainty, as we are unable to monitor photometric variability due to atmospheric conditions over the course of the observation, which represents the dominant source of uncertainty for the stellar photometry.
To measure the uncertainty on the planet photometry we take the standard deviation of the residuals in an annulus at the separation of the companion, masking out the planet itself.

The histograms of Figure \ref{fig:ifuframes} show that the assumption of Gaussian errors across the entire frame is inconsistent with the noise, and would underestimate the tails of the distribution.
\cite{parait_2019_stim} demonstrate that a Laplacian provides a better fit to the tails of the residual distribution than a Gaussian, while \cite{mawet_2014_limits} shows that the residuals tend to follow a Student-t distribution. 
We find that a Student-t distribution best matches the full frame residuals. 
However, as the likelihood function for a general Student-t distribution is not analytic, and a Gaussian distribution accurately captures the residuals to $2.15\sigma$, we will continue to follow the standard practice of defining uncertainties as the Gaussian standard deviation.
Taking \verb|KLIP| as an example, the best fit Student-t distribution has 1.75 DoF, a mean of $2.23\times10^{-8}$ and a width $t=5.29\times10^{7}$. 
At the point where the Gaussian distribution intersects the Student-t distribution, 89\% of the residuals are enclosed, compared to 97\% if the residuals were Gaussian distributed.
We also note that, relative to the speckle subtraction algorithms, \verb|ANDROMEDA| shows an excess of 10$\sigma$ outliers in the residuals, leading to difficulties in distinguishing between true positives and false positives, consistent with the findings of \cite{CantalloubeEIDC}. 
The long tails of these distribution add additional noise to each frame and need to be accounted for for the \textit{detection} of planet candidates in order to avoid false positives. 
However, for a known companion where we are concerned with inferring physical parameters to within 1-2$\sigma$ confidence intervals, accounting for the 90\% of the noise that is contained within the Gaussian fit to the residuals is sufficient for defining the uncertainties.

Thus the total uncorrelated uncertainty for the $i^{th}$ wavelength channel is given as:
\begin{equation}\label{eqn:errors}
    \mathbf{C}_{ii} = \sigma_{i,star}^{2}\left(\frac{f_{i,pl}}{f_{i,star}}\right)^{2} + \sigma_{i,residual}^{2}.
\end{equation}
The method described here provides an empirical estimate of the covariance of the noise after high-contrast image processing.
However, as it relies on measurements of mean pixel intensities in a residual image, it is only applicable for speckle-subtraction methods.
As \verb|ANDROMEDA| produces an estimate of planet contrast at each pixel location, rather than residual noise following PSF subtraction, this method cannot be directly applied to the processed \verb|ANDROMEDA| frames.
An example of such a frame is shown in Figure \ref{fig:ifuframes}, where highly structured noise is visible in the frame. 
The noise pattern is highly correlated through wavelength space, and indeed would lead to very strong residual correlation.
Rather than applying the procedure for measuring the covariance matrix for \verb|ANDROMEDA|, we instead rely on the estimate of the standard deviation that is also provided by the algorithm, that is also measured during the likelihood minimization.

Figure \ref{fig:correlations} shows the results of computing the correlation matrix for both the {\tt KLIP} and {\tt PynPoint} reductions. 
There is a strong, narrow correlation component along the diagonal with a width of around 2-3 pixels, with a weaker correlation extending out to 10 pixels in width. 
In the SPHERE data, the correlation decreases in the water absorption features at 1.15 $\upmu$m and 1.4 $\upmu$m. 
The {\tt KLIP} data typically displays stronger correlations than the {\tt PynPoint} reductions.
This difference may be because the {\tt pyKLIP} implementation of the {\tt KLIP} algorithm uses only the most correlated frames from the PSF library to build the PSF model, which introduces an additional source of correlation in the data.

\subsection{Bias Correction}
PCA methods tend to see increased self-subtraction as the number of principal components increases, naturally leading to poorer extractions at as the number of components increases \cite{Lagrange2010_bias}
Therefore we also considered an empirical estimate of the uncertainty by injecting and recovering a sample of planets in an annulus at the location of the planet. 
The standard deviation of the recovered spectra provide a measure for the uncertainty due to planet position and variation in the effectiveness of the post-processing. 
This can also be used as a method to correct for bias introduced by self-subtraction caused by the post-processing algorithm: by comparing the recovered spectra to the known input, a scaling factor can be computed.
This can then be used to mitigate the self-subtraction induced by the PCA processing \citealt{Marois2014Bias,Gerard2016Bias,Ruffio2017FMMF}.
We found that when applying such a bias correction, the $\chi^{2}$ between the injected and recovered spectra was often worse than that of the nominal spectral extraction.
As the noise properties are not truly azimuthally symmetric, the average bias correction does not provide a good correction for any individual planet location, particularly at the relatively faint contrasts considered in this work.
Any improvement in the spectral extraction is dominated by the both the changes to the shape of the spectrum introduced by the data processing and the random noise of the measurement. 
Therefore we choose not to include bias correction as a step in our post processing, and do not include uncertainty from injection and recovery tests in our error estimate.
The finding that bias correction can reduce the accuracy of the spectral extraction is surprising, and warrants further investigation into where this widely-used technique should be applied. 
We leave such a study to future work.

\begin{figure}[t]
    \centering
    \includegraphics[width = \linewidth]{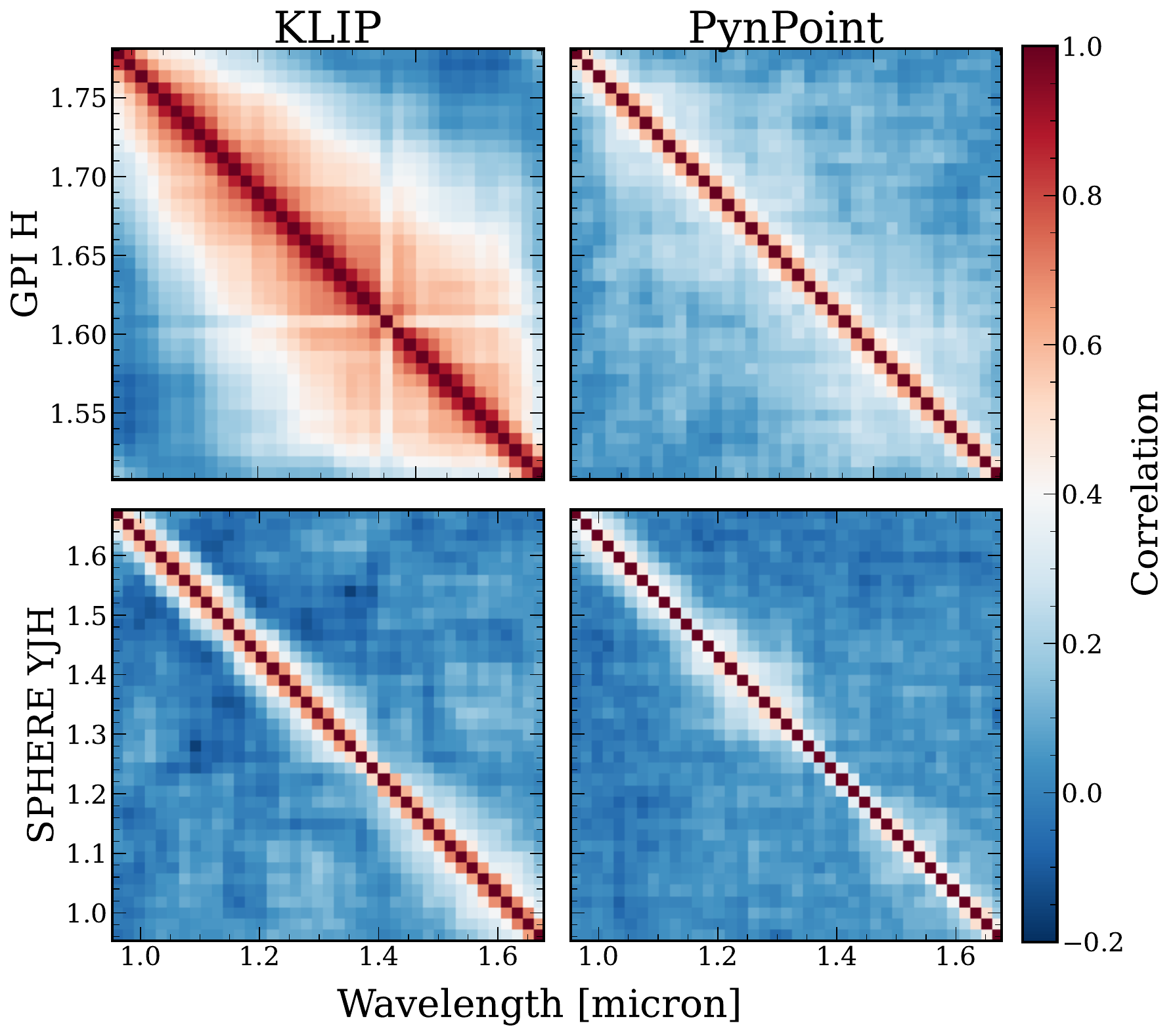}
    \caption{Correlation matrices for each dataset for HR~8799~e, with the GPI H-band data shown in the top row and the SPHERE YJH data on the bottom. Following the processing using {\texttt KLIP} (left) or {\texttt PynPoint} (right), we calculate the correlation and covariance matrices as described in Section \ref{sec:covar}. The correlation is computed as in Equation \ref{eqn:cor}. 
    The GPI data is more strongly correlated than the SPHERE data, particularly following the {\texttt KLIP} processing. 
    The SPHERE data shows structure similar to the correlation matrix, with the correlation width following the shape of the water absorption spectrum.}
    \label{fig:correlations}
\end{figure}

\subsection{Impact of Covariance on Retrieved Parameters}\label{sec:covimpact}
\begin{figure*}[t]
    \centering
    \begin{subfigure}{0.65\textwidth}
    \includegraphics[width=\linewidth]{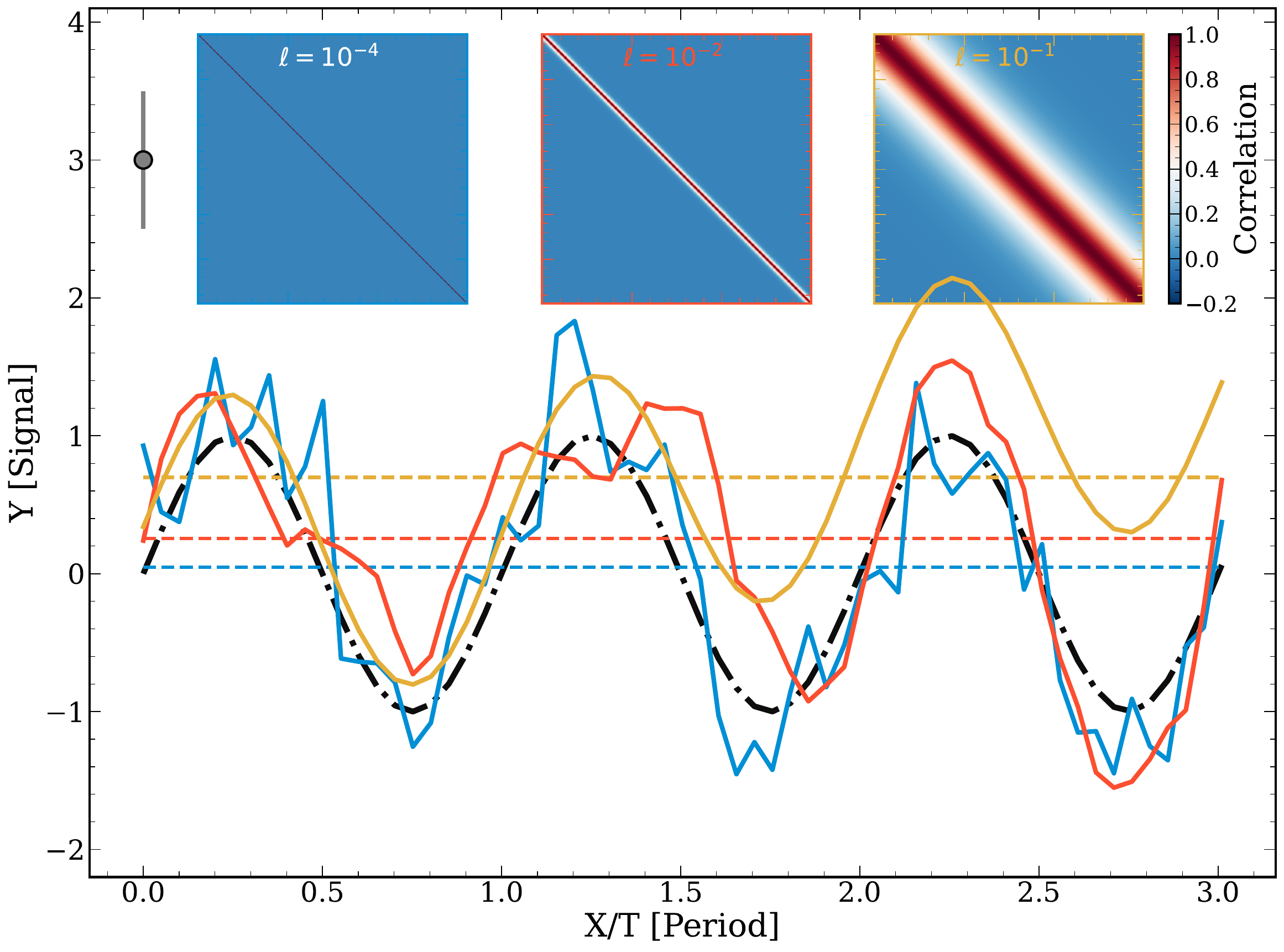}
    \end{subfigure}
    \begin{subfigure}{0.328\textwidth}
    \includegraphics[width=\linewidth]{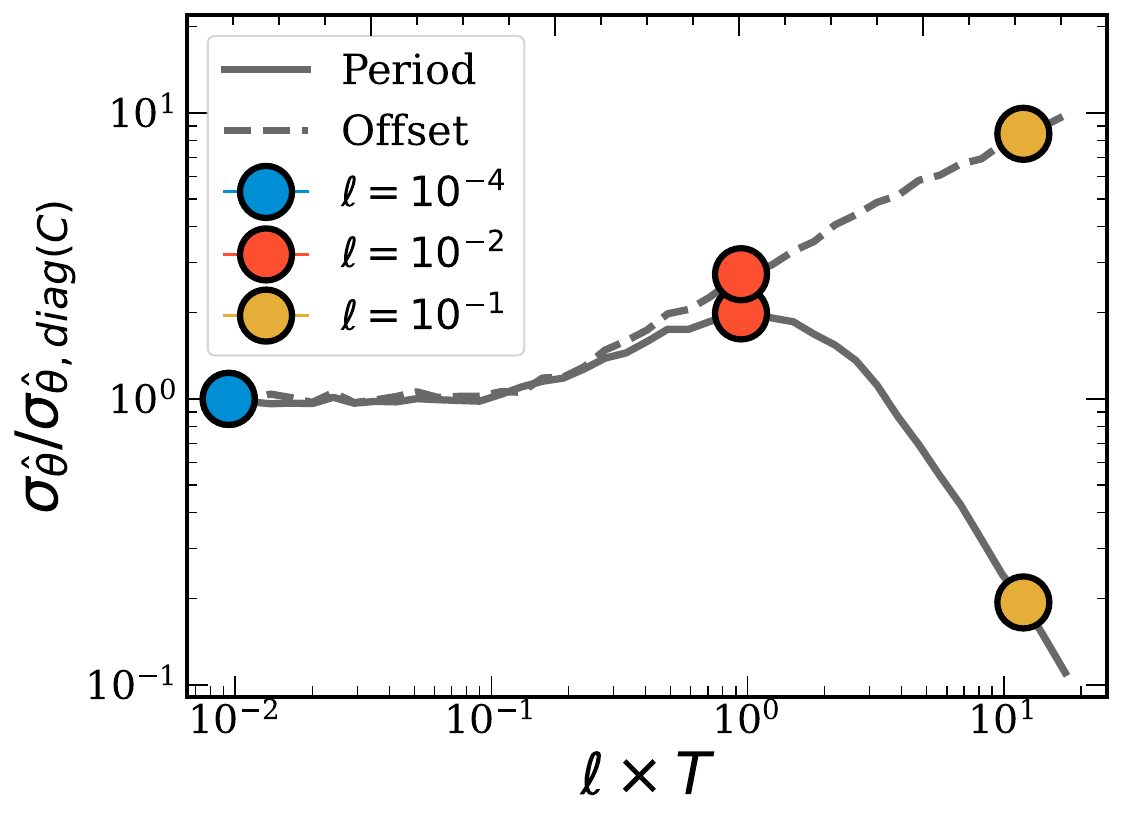}\\
    \includegraphics[width=\linewidth]{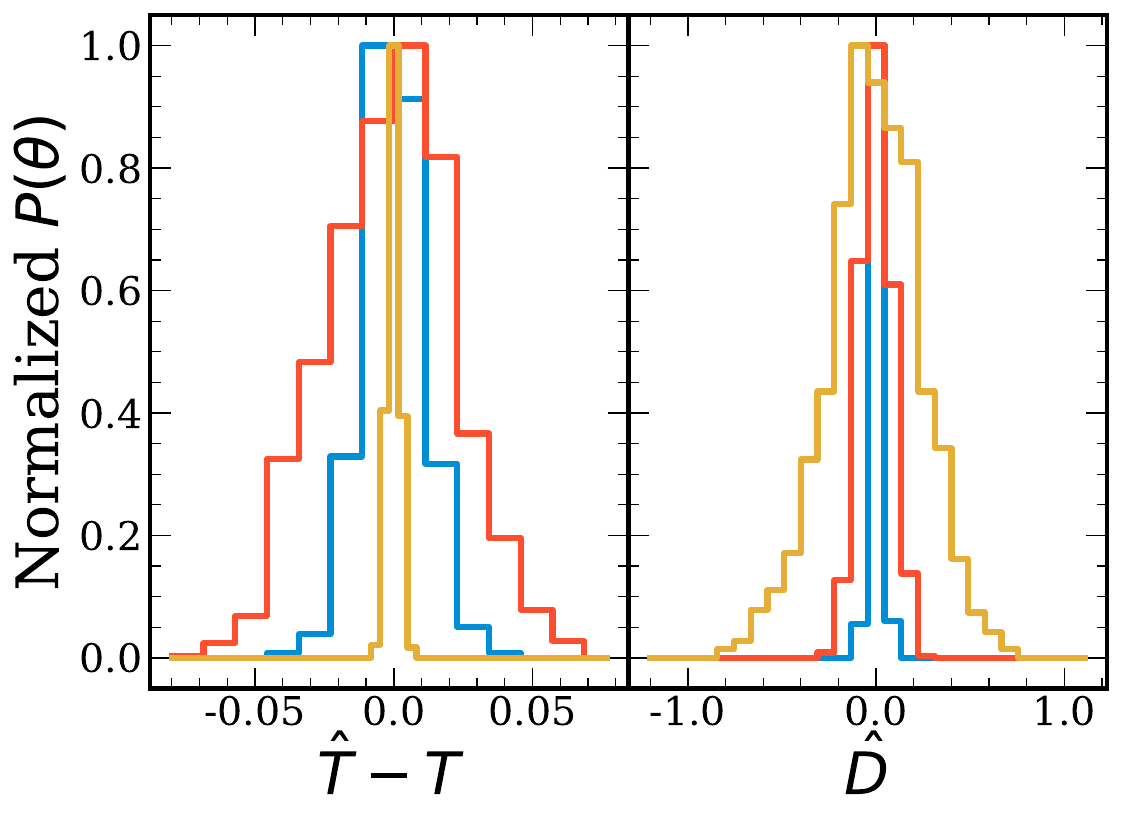}
    \end{subfigure}

    \caption{\textbf{Left:} Data drawn from a toy sinusoidal model (Eqn. \ref{eqn:sinmodel}) when considering three cases of covariance in the data. The first (blue) is data drawn from a univariate Gaussian distribution with no correlations, as shown in the left inset. The second (orange) is drawn from a multivariate Gaussian distribution where the correlation length scale - as defined by the $\ell$ parameter of the Mat\'{e}rn kernel (Eqn. \ref{eqn:matern}) - is less than the period of the model (center inset). The third (yellow) is drawn from a multivariate Gaussian distribution with a correlation length scale greater than the period of the model (right inset). In the background the true input is plotted in gray. 
    The gray datapoint indicates the 1$\sigma$ error bar associated with each data point.
    \textbf{Right:} In the top panel, the posterior width of the period (solid gray line) and offset (dashed gray line) parameters scaled to the uncorrelated case as a function of the ratio between the correlation length scale and the period of the sine (so the model length scale). Marked in blue, orange and yellow are the draws plotted in the right panel. The histograms in the bottom panel are the posterior histograms for the period (left) and offset (right) for each of the highlighted cases.}
    \label{fig:CorrelationScale}
\end{figure*}
In the previous section we discuss how to measure a covariance matrix for IFS data.
Here we explore how the covariance impacts Bayesian inference, that is, when estimating the parameters of the model used to explain the observations. 
\cite{greco_brandt_2016} demonstrated how failing to include the full covariance matrix when fitting atmospheric models to IFS data will result in overly confident and biased parameter estimates.
In \ref{sec:injrets} we will expand their work to higher dimensional models using atmospheric retrievals, but first we want to pedagogically understand how the covariance is linked to the posterior probability distributions.
We will show how the precision of a posterior parameter estimate depends on the ratio between the length scale of the correlation in the data and the length scale over which the parameter introduces changes in the model spectra. 
If this ratio is larger than unity, the posterior width will decrease relative to the case without correlation.  
If the ratio is about unity the posterior width will increase. 

Consider a toy model, where the data $\mathbf{y}$ is given by a simple sine function with period $T$ and offset $D$:
\begin{equation}\label{eqn:sinmodel}
    y_{i} = \sin\left(2\pi x_{i}/T\right) + D.
\end{equation}
In the context of atmospheric parameters, we can think of this model as the first term of a Fourier series, which can be used to describe an atmospheric spectrum to arbitrary precision if extended to a high enough order. 
While the variation in the spectrum due to physical parameters is more complicated than this model, we can view an offset in the toy model as a change in the overall flux, while a change in the period would be reflected in the spectral shape, such as the near infrared water features.

The period and offset were arbitrarily chosen to be 30 and 0 respectively. 
Assuming that this toy model describes an observed experimental setup, we construct a synthetic data set containing a total of 300 points, with $x$-coordinate values from  1 to 300, that is, 10 periods.
We will apply different noise models to these toy data, and use nested sampling as implemented in \verb|MultiNest| \citep{feroz_multimodal_2008,feroz_multinest_2009, feroz_importance_2013} to retrieve the value of the period parameters $T$ and $D$. 
Such nested sampling algorithms improve on MCMC techniques to sample large parameter spaces efficiently, gradually restricting the sampling volume to regions of high likelihood.
They are also robust to multimodal posterior probability distributions, and provide an estimate for the Bayesian evidence, $\mathcal{Z}$, as well as the posterior distributions and maximum likelihood fit.

We use the Mat\'{e}rn 3/2 kernel \citep{Rasmussen2006_GPsML} to describe the covariance of our dataset:
\begin{equation}\label{eqn:matern}
    C_{3/2}\left(d\right)_{ij} = \sigma_{i}\sigma_{j}\left(1 + \frac{\sqrt{3}d_{ij}}{\ell}\right)\exp\left(\frac{-\sqrt{3}d_{ij}}{\ell}\right).
\end{equation}
This is a model of for the correlation between two points separated by distance $d_{ij}$, where we can adjust characteristic correlation length scale through the correlation length parameter, $\ell$. 
As $\ell$ decreases, the correlation matrix becomes more diagonal, while as $\ell$ increases the data becomes more strongly correlated across broad scales.
The strength of the correlation is determined by the uncertainty on each point ($\sigma_{i}$), which we set to a constant value of 0.5 for each data point.

In Figure \ref{fig:CorrelationScale}, we show three instances of the correlation matrix for $\ell = 10^{-4},10^{-2}$ and $10^{-1}$ , ranging from uncorrelated to strongly correlated noise.
Plotted in gray is the noise-less data, and the coloured lines show noise instances drawn from each of the three correlation matrices.
For the diagonal case ($\ell=10^{-4}$), we see that the data are randomly scattered around the true model.
With no correlation, we see true, univariate Gaussian noise.
As the correlation scale increases to $\ell=10^{-2}$, we see that the data appear smoother, and the variations occur on larger spatial scales than in the case without any correlation.
Finally, with $\ell=10^{-1}$, we see that the data are offset from the ground truth, but do not have any small-scale scatter.
This is the impact of covariance on the data: as the correlation length increases, every point is more strongly determined by the initial random draw any other point (effectively there are less points, as there are less independent measurements). 
In summary, we observe high frequency variation due to noise if the correlation length is small. In this case the mean of the data, parameterized by parameter $D$ should be accurately and precisely inferred. 
As the correlation length increases, the scatter of the mean $D$ across multiple draws increases, but we see less small-scale variation, allowing a better estimate of the period, $T$.

We note here that the Mat\'{e}rn 3/2 kernel is only one model for the covariance, and is not perfectly suited for IFU data. 
A more robust model (such as described in \citealt{greco_brandt_2016}) would incorporate both a broad correlation term and a diagonal Gaussian term to the correlation matrix, which would introduce small-scale scatter in the data, even with a large correlation scale. 
Nevertheless, this is a suitable toy model to explore how changing the correlation length scale impacts parameter estimation.

To determine the impact on parameter inference, we vary the correlation matrix across a range of $\ell$ values from $10^{-4}$ to $10^{0}$, and use \verb|pyMultiNest| \citep{buchner_x-ray_2014} with 400 live points to fit the true model, accounting for the covariance in the likelihood. We did not perturb the toy data set by with an error model as defined by the covariance matrix, so we run noise-free retrievals.
This is equivalent to running multiple inferences where the data are perturbed by draws from the covariance matrix and averaging over the posteriors of each inference.
We set uniform priors on the period $P(\omega) = \mathcal{U}(0,100)$ and offset $P(D) = \mathcal{U}(-10,10)$.
The results are not sensitive to the choice of number of live points $(n_{\rm live} >> n_{\rm param})$ or priors.
The upper right panel of Figure \ref{fig:CorrelationScale} shows the ratio between the width of the posterior distribution for this parameter and the width of the distribution in the case of univariate Gaussian noise (i.e., no correlation). 
We observe that as the correlation length scale approaches the length scale of the sine function (the period) the width of the posterior increases: the correlation introduces variations in the data on the scale of the period, making it difficult to estimate the parameter.
This is the effect described in \cite{greco_brandt_2016}, where accounting for the covariance matrix when fitting atmospheric models will increase the posterior width.
However, as the correlation scale continues to increase to scales larger than the period, we see that the posterior width decreases to values lower than in the case of uncorrelated noise. 
As is visible in the data in the left panel, without small scale variations to introduce uncertainty in the period, it becomes easier to estimate this parameter, at the cost of increased uncertainty in the estimate of the offset parameter $D$.

\subsubsection*{Effects of ignoring covariance}
It is often the case that the full covariance matrix is not used when performing atmospheric retrievals, and we wanted to explore the impact of using only the diagonal terms when fitting a model to correlated data. 
Figure \ref{fig:chi2vcorrelaiton} shows the best-fit reduced $\chi^{2}$ as a function of the ratio between the period and the correlation length scale, as in the right panel of Figure \ref{fig:CorrelationScale}. For each $\ell$, we perform an ensemble of 25 retrievals using Multinest in order to reduce the scatter and to measure the uncertainty in the $\chi^{2}$ due to the variation between individual noise instances.
In this case, the data are perturbed by draws from the covariance matrix, in order to test the impact of using the incorrect covariance in the likelihood when the data are correlated.
We define $\nu$ as the number of data points (300) minus the number of parameters (2). 
This procedure is repeated using both the full covariance matrix, $\mathbf{C}$, in the likelihood, as well as using only the diagonal elements of the matrix - i.e. assuming that the data are uncorrelated.
We find that the reduced $\chi^{2}$ is a useful metric if the covariance is properly accounted for. If the data are correlated and an only the uncorrelated uncertainties are used in the likelihood then the reduced  $\chi^{2}$ will be underestimated, and the scatter of the  $\chi^{2}$ increased.
Often a  $\chi^{2}/\nu<1$ is interpreted either as overfitting of the data or overestimation of the uncertainties.
However, we demonstrate here that for $n_{\rm param} < n_{\rm data}$ a $\chi^{2}/\nu<1$ can be interpreted as an \textit{underestimation} of the correlation of the data.
\begin{figure}[t]
    \centering
    \includegraphics[width=\linewidth]{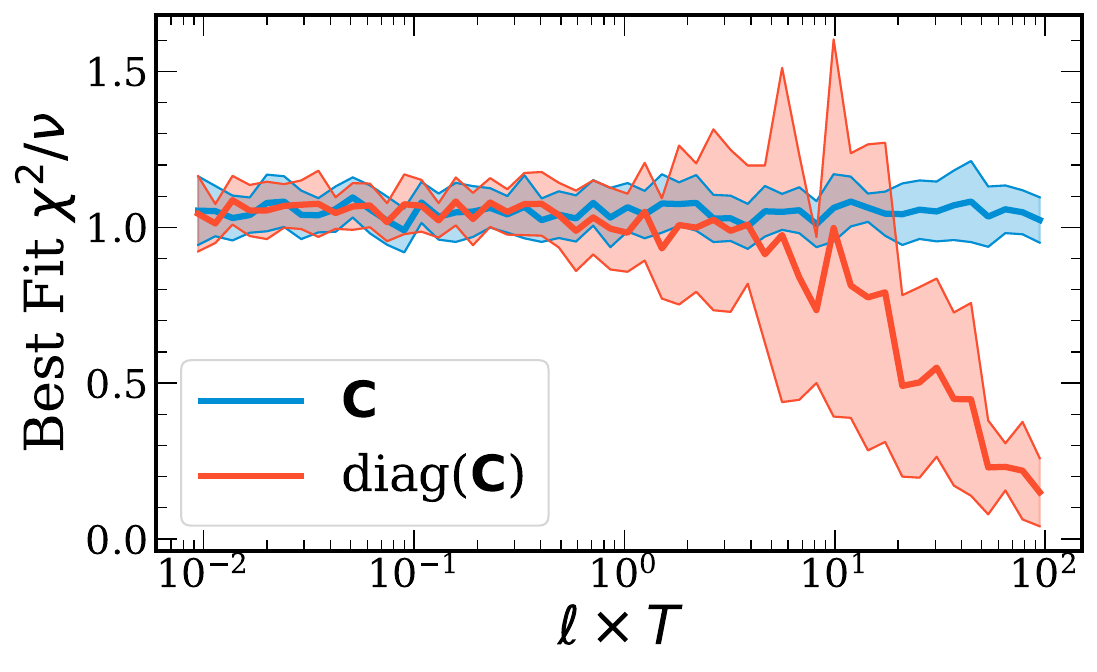}
    \caption{The best fit $\chi^{2}/\nu$ as a function of the ratio between the correlation length scale (proportional to 1/$\ell$) and the period, $T$. The $\chi^{2}$ was computed for fits of equation \ref{eqn:sinmodel} to data perturbed by draws from the covariance matrix, varying the correlation length scale. For each $\ell$, 25 Multinest retrievals were run in order to compute the uncertainty on the $\chi^{2}$, shown as the shaded region around the mean.  In blue, the covariance is properly accounted for in the likelihood, while in orange only the diagonal of the covariance is used in the likelihood. In order for the reduced $\chi^{2}$ to be a useful metric, the covariance must be properly accounted for.}
    \label{fig:chi2vcorrelaiton}
\end{figure}

\subsubsection*{Applicability to atmospheric retrievals}
We expect similar effects to be present in atmospheric retrievals with correlated data.
Parameters that affect model spectra on wavelength scales smaller than the correlation scale may be retrieved to higher precision than expected if the uncertainties were uncorrelated, while parameters that are sensitive at approximately the correlation length scale will have larger posterior uncertainties.
As seen in Figure \ref{fig:correlations}, the correlation length can be a appreciable fraction of the total data, particularly in the case of the KLIP reduction of GPI data.
Large scale correlations in the data can introduce offsets in the average flux measurement, which can lead to inconsistencies between datasets from different instruments or measured during different epochs.
Correlations on moderate scales can alter the spectral shape, in turn impacting parameter estimates.
For example, the surface gravity is particularly sensitive to the shape of the H-band, and changes to the shape of this band will lead to biased estimates.
Thus for IFS data it is critical to account for the covariance matrix when fitting models to the data, in order to correctly capture the noise structure imprinted onto the signal.

\begin{figure*}[t!]
    \vspace{0.2cm}
    \centering
    \hspace{1cm}$\boldsymbol\chi^{\boldsymbol2}\mathbf{/n_{\rm data}}$
    \hspace{8cm}$\mathbf{1 - \frac{1}{N}\left|\sum_{i}^{N}\frac{f_{i}}{y_{i}}\right|}$

    \includegraphics[width=0.48\linewidth]{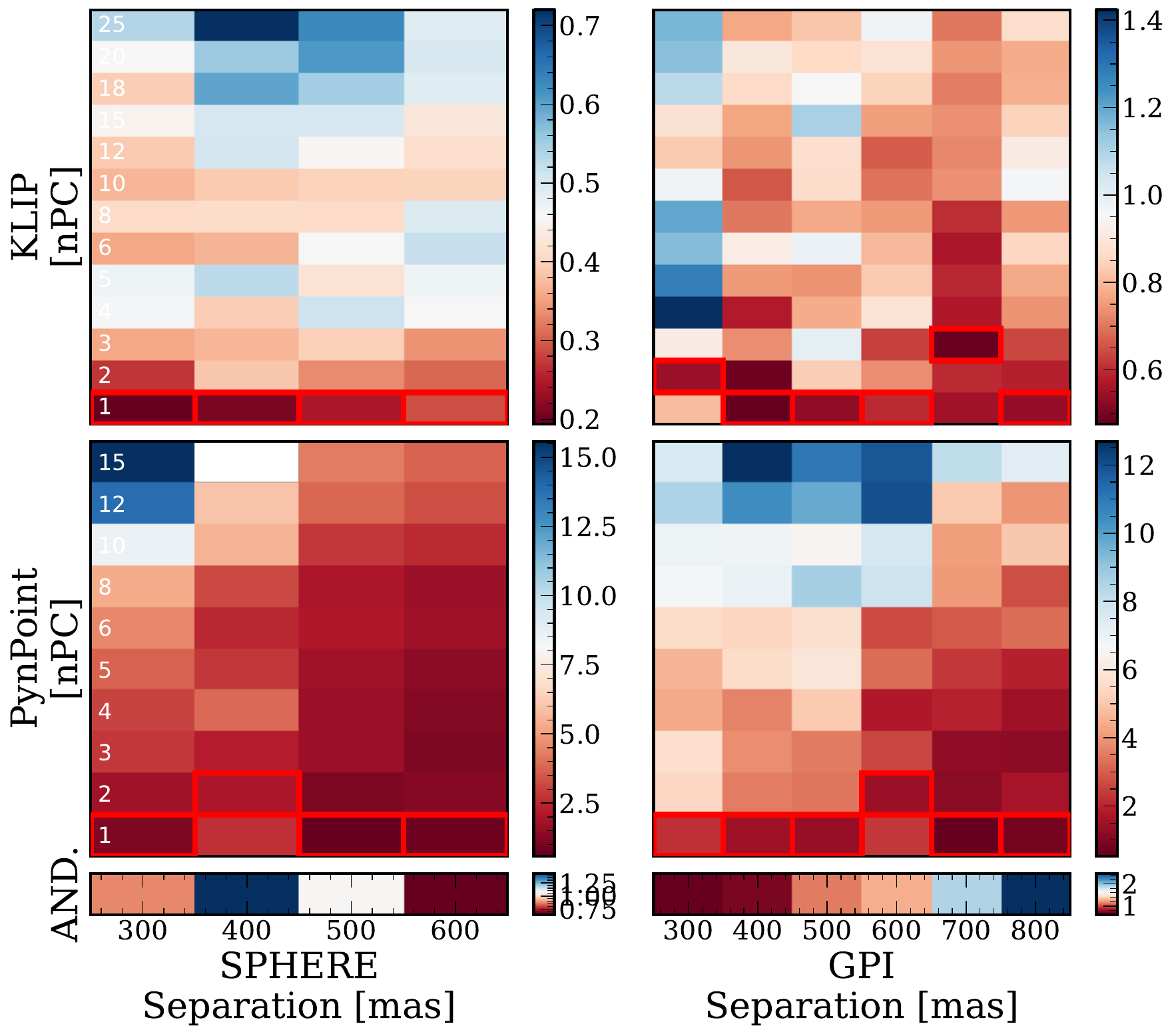}
    \includegraphics[width=0.48\linewidth]{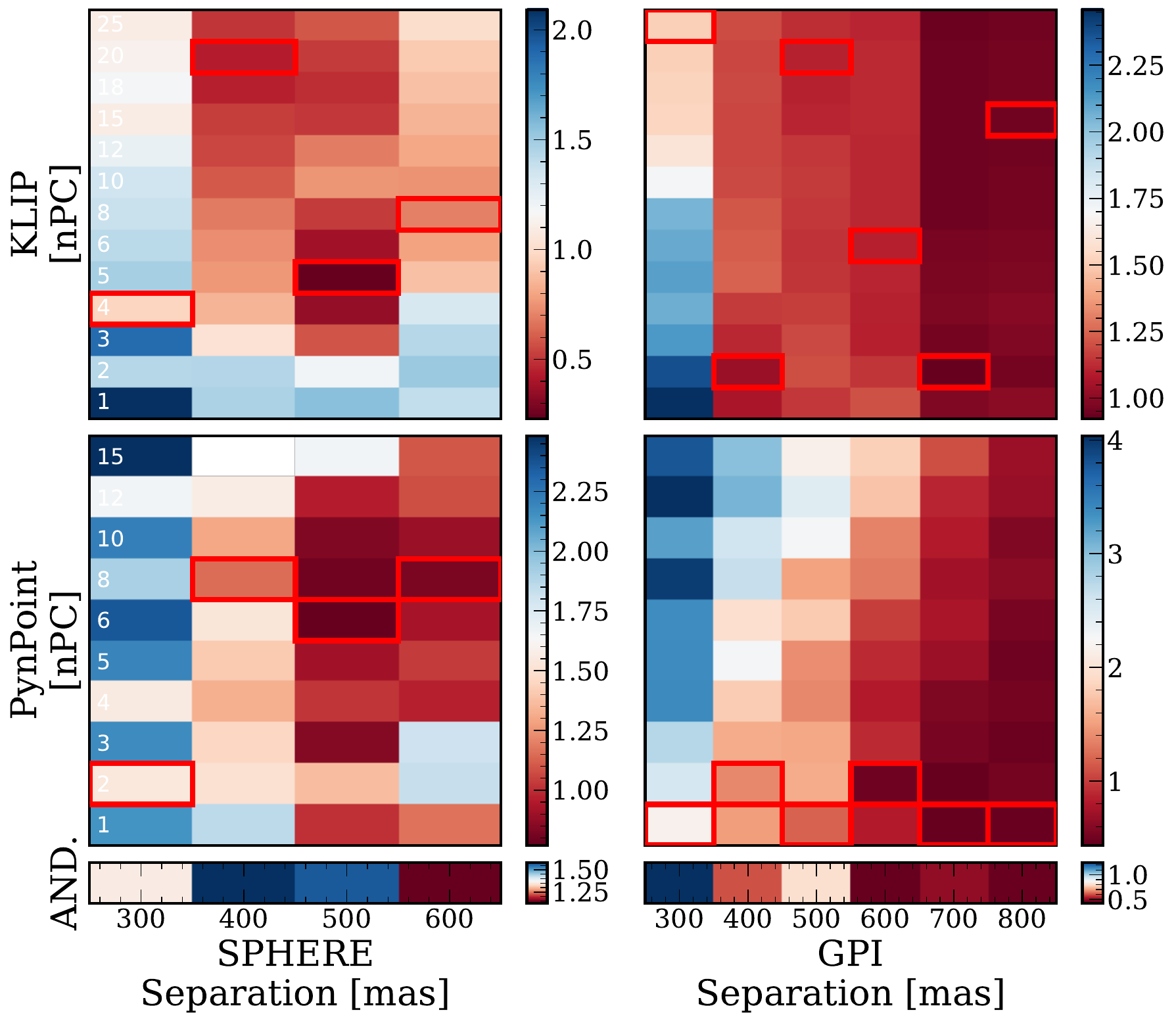}

    \caption{\label{fig:PCAAlgoComp} \textbf{Left:} $\chi^{2}$ values mapped across position and number of principal components for each algorithm. 
    The colour scale indicates median $\chi^{2}/n_{\rm data}$  from the injections at 4 separate position angles. The range of the colour scale for each sub plot is different in order to capture the variation within a single map. Highlighted in red are the optimal extractions for each separation. \textbf{Right:} The same as the left panel, but calculated using the relative discrepancy instead of the $\chi^{2}$.}
    \label{fig:Chi2SeparationPC}
\end{figure*}

\section{Injection Testing}\label{sec:Injections}
In order to best extract a true signal, we want to optimise the data-processing parameters.
However, without knowledge of the ground truth spectrum, it is unclear how these parameters should be tuned a priori.
By injecting fake companions with a known spectrum, applying the post-processing, and comparing the extracted spectrum to the input we can then optimise the parameters, and use this setup to extract the true planet signal.
In particular, we try to optimise the choice of the number of principal components used in PSF subtraction for {\tt KLIP} and {\tt PynPoint} as a function of the separation.
This injection-extraction study will also provide us with a metric for comparing the three algorithms described in Section \ref{sec:algs}.

Using the {\tt pyKLIP} injection tool we injected companions into both the SPHERE and GPI HR 8799 datasets.
The normalised stellar PSF was used both as a model for the planet PSF and to scale the bulk contrast of the injected companion.
The spectrum was convolved with a gaussian kernel to the instrumental resolving power, and binned to the instrumental wavelength grid using the {\tt rebin_give_width} function available in petitRADTRANS, which accounts for non-uniform bin sizes as the number of pixels per instrumental resolution element varies with wavelength.
Only a single planet was injected at a time before the data processing, which was repeated for each planet position in order to avoid potential contamination from nearby signals.
We injected the companions at varying positions into both the SPHERE and GPI datasets, with a spectrum generated using \verb|petitRADTRANS| as described in Section \ref{sec:retrievals}.
These were positions representative of the known separations of the inner three companions. 
Planets were injected at position angles from 120\textdegree\ to 240\textdegree\ from the location of HR~8799~e in 30\textdegree\ increments, and between 300 -- 800 mas in 100 mas steps. 
This process was repeated for the SPHERE YJH and the GPI H-band datasets at mean contrasts from $10^{-7}$ to $10^{-4}$.

Once the data were prepared, we ran each of the three data processing algorithms on each injected dataset, spanning a range of algorithmic parameters. While we could not exhaustively study the effect of each parameter, we chose to focus on the impact of the number of principal components used during PSF subtraction in order to optimize the spectral extraction.
Other parameters, such as the \verb|flux_overlap| parameter in 
\verb|KLIP|, or the filtering fraction in \verb|ANDROMEDA| were set based on suggested values from previous studies \citep{zurlo_first_2016, Cantalloube2015} or from qualitative examination of the post-processed data. 
Several parameters, such as the \verb|tolerance| and \verb|merit| parameters of \verb|PynPoint| were chosen to ensure accurate extractions within reasonable computation time.
Various geometric parameters, such as the inner and outer working angles together with the width parameter in \verb|ANDROMEDA|, or the \verb|subsection| and \verb|annuli| parameters of \verb|KLIP| were set based on recommendations from the documentation\footnote{\url{https://pyklip.readthedocs.io/en/latest/fm_spect.html}}, and ensuring that the region under consideration would contain the entirety of the planet signal, extending to at least twice the FWHM of the signal.
A full table of parameter choices for each algorithm is given in Table \ref{tab:algoparams}.

\subsection{Choice of goodness-of-fit metric}
We considered several goodness-of-fit metrics with which to determine the optimal extraction, including the signal-to-noise ratio ($S/N$), the relative discrepancy ($e$) and reduced $\chi^{2}$ ($\chi^{2}/n_{\rm data}$).
We take the median of each metric across the five different position angles where planets were injected.
We exclude spectra that are over 20$\sigma$ discrepant from the input, or that display strong outliers with contrast $>2\times10^{-5}$, though the results are robust to including the outlying data. 
Each of these metrics identified different optimal spectra.

The mean $S/N$ always identified the spectra processed using the largest number of components as optimal. 
However, the resulting spectra do not correctly retrieve the shape of the input spectrum, because they typically overestimate the flux, so we did not consider this metric further.

We define the mean relative discrepancy $e$ between a measured flux $\vec{s}$ and known input spectrum $\bar{\vec{s}}$ as 
\begin{equation}\label{eqn:reldist}
    e = 1 - \frac{1}{N}\left|\sum_{i}^{N}\frac{s_{i}}{\bar{s}_{i}}\right|.
\end{equation}
To identify the best fit spectrum we simply find the minimum value of this function.
In contrast to the $\chi^{2}$ or other distance metrics, the discrepancy is is invariant of the magnitude of the measured quantity, and so provides a metric to compare spectra injected at different contrasts.

The $\chi^{2}$ is a standard metric for measuring the similarity of distributions, but can also favour measurements with overestimated uncertainties.
The $\chi^{2}$ value between the extracted spectrum $\mathbf{s}$ with covariance $\mathbf{C}$ and the known injected spectrum $\mathbf{\bar{s}}$ was calculated for each post-processed dataset as
\begin{equation}\label{eqn:chi2}
    \chi^{2} = \left(\mathbf{s}-\mathbf{\bar{s}}\right)^{T}\mathbf{C}^{-1}\left(\mathbf{s}-\mathbf{\bar{s}}\right).
\end{equation}
We present $\chi^{2}/n_{\rm data}$, dividing by the number of wavelength channels $n_{\rm data}$ to allow for a more straightforward comparison between instruments. 
We do not subtract the degrees of freedom from the number of data points as is typical when computing the reduced $\chi^{2}$, as principal components are not free parameters in a statistical sense, thus making the definition of degrees of freedom challenging.

\begin{figure}[t]
    \centering
    \includegraphics[width=\linewidth]{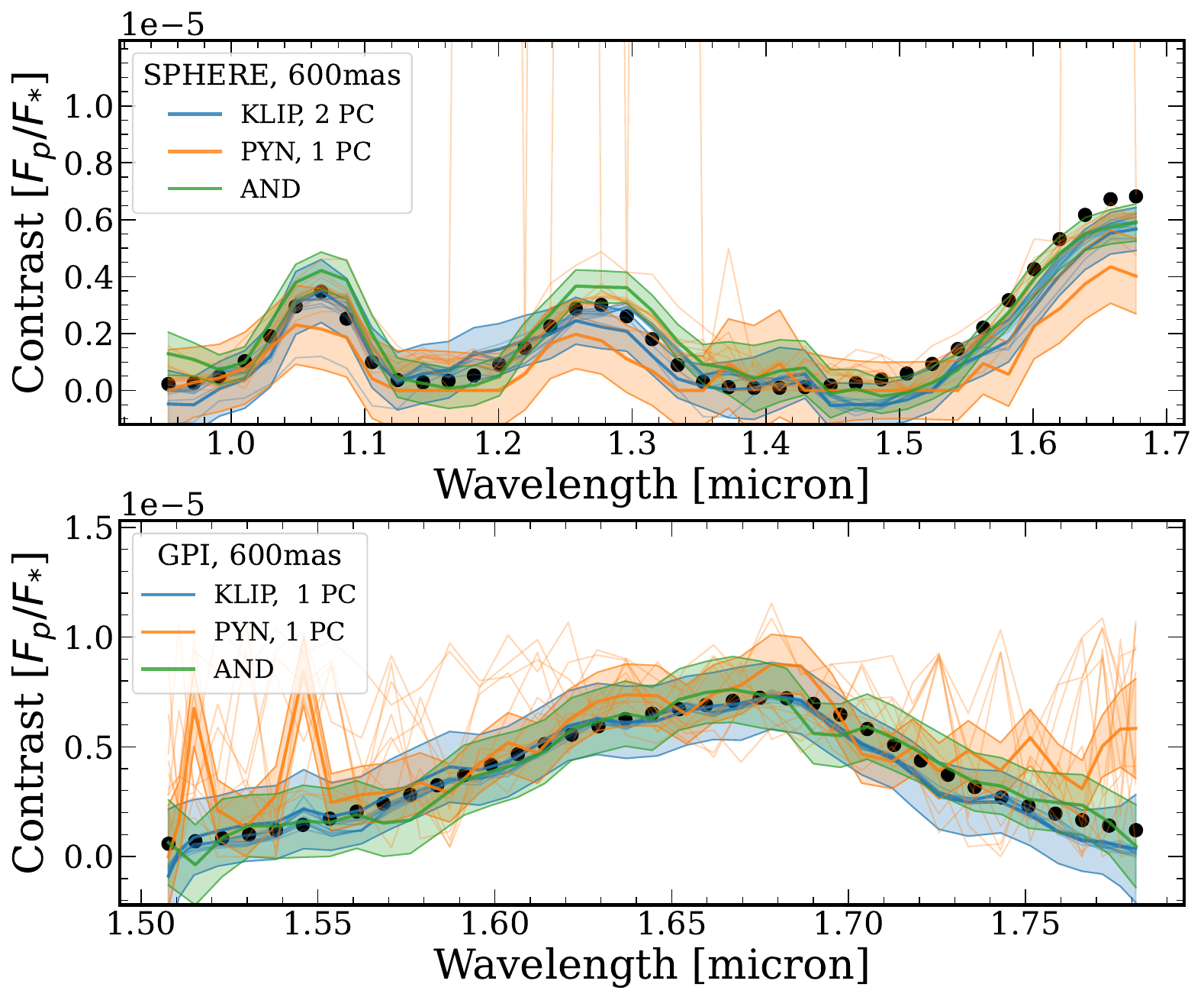}
    \caption{Typical spectral extractions for injected planets located at separations of 600 mas. 
    These spectra are representative of the HR~8799 planets with ($F_{p}/F_{*}\sim2\times10^{-6}$).
    The injections into the SPHERE data are shown on the top panel, the GPI on the bottom.
    Each injected planet was positioned 150\textdegree\ from HR~8799~e.
    Extractions for each algorithm are plotted, with the best fit spectrum ($\chi^{2}$) and 1$\sigma$ error bars from the diagonal of the covariance matrix highlighted by the shaded region.
    The faint lines show the variation in the extractions using different numbers of principal components.}
    \label{fig:InjectionExtraction}
\end{figure}

\begin{figure}[t]
    \centering
    \includegraphics[width=1\linewidth]{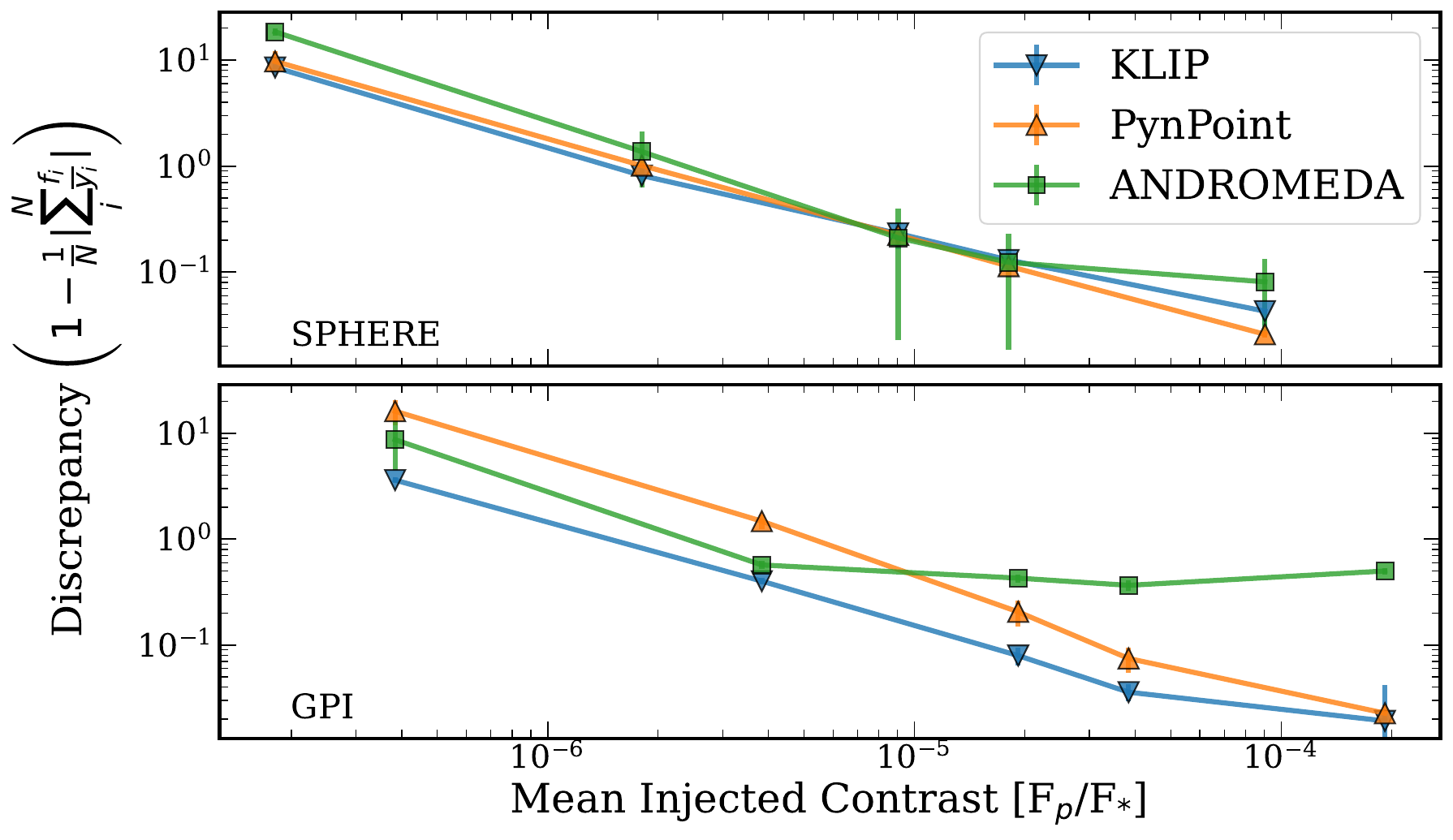}
    \caption{Best-fit discrepancy (eqn. \ref{eqn:reldist})  as a function of input contrast at 400 mas.  The top panel shows the results for injections into the SPHERE data cube, while the bottom is for GPI. The injections were repeated at three position angles, and the uncertainty presented is the standard deviation of these measurements.}
    \label{fig:Chi2Contrast}
\end{figure}

\subsection{Optimising Spectral Extractions}
We present in Figure \ref{fig:Chi2SeparationPC} the map of $\chi^{2}/n_{\rm data}$ (left) and mean relative discrepancy (right) as a function of both separation and number of principal components used, taking the median across the injections at different parallactic angles.
Extracted spectra for typical injected fake companions are shown in Figure \ref{fig:InjectionExtraction}.
The precision and accuracy of our spectral measurements depends strongly on the separation, visible in the variation of the both metrics.
There is strong position-dependent variation in the shape of the extracted spectra, including cases where the injected companion is not detected in any wavelength channel, as well as cases where the peak contrast is overestimated by a factor of 2.
Examining Figure \ref{fig:ifuframes}, we see that beyond about 400 mas from the host star the noise properties are relatively unstructured, while at 400 mas and closer the GPI data is dominated by the residual speckle noise.
Such a trend is also present in the SPHERE data, though the speckle dominated regime extends out to only 300 mas.
This transition in the underlying noise properties together with the greater angular displacement at wider separations results in the improved detections at wider separations.
This can be disguised by the $\chi^{2}$ metric, where large uncertainty estimates at small separations can result in a better $\chi^{2}/n_{\rm data}$, while the mean relative discrepancy provides a clearer trend as a function of separation.
When using $\chi^{2}/n_{\rm data}$ as the goodness-of-fit metric, we find that both {\tt KLIP} and {\tt PynPoint}  favour low numbers of principal components.

Such results depend strongly on both the dataset, the reduction used, and the choice of metric.
The SPHERE extractions are universally better than the GPI, largely due to the brightness of the injected spectra in the Y and J bands, the smaller inner working angle and pixel scale, and the longer integration time.
{\tt PynPoint} most strongly favours low numbers of principal components, across both metrics and at all separations.
However, when we consider the relative discrepancy between the input and extracted spectra, and find that the number of PCs components favoured is much higher than when using the $\chi^{2}$, particularly for {\tt KLIP}.
These spectra may more closely match the shape of the input spectra, but may also underestimate the uncertainties, leading to them being disfavoured by the $\chi^{2}$.
While the shape of the spectrum does depend on the number of components used, it is more strongly dependent on the particular location in the frame where it is injected.

Each algorithm displayed its own trends in the quality of its spectral extractions.
\verb|KLIP| produced smooth spectra, with systematics that were relatively consistent in shape over the full range of principal components used. 
However, it struggled to recover the brightest sources accurately, reproducing the over-subtraction effect described in \cite{pueyo_klip}.
In contrast, \verb|PynPoint| and \verb|ANDROMEDA| performed worse at fainter contrasts, as demonstrated in Figure \ref{fig:Chi2Contrast}.
The \verb|PynPoint| spectra are more dominated by random scatter than by systematic variation, reflected in the typically diagonal correlation matrices.
\verb|ANDROMEDA| produced some of the best overall fits, but struggled to achieve the correct flux calibration, both over- and under-estimating the flux in different cases. 
For the bright injection case, \verb|PynPoint| consistently performed the best, producing the lowest $\chi^{2}$ values for each dataset and separation.
\verb|KLIP| struggled to extract the brightest spectra, over-subtracting the planet signal at the red end of the SPHERE data.
This effect was more severe when larger numbers of principal components were used.
However, \verb|KLIP| also displayed a tendency to over-estimate the flux of the signals injected into the GPI data.
\verb|ANDROMEDA| was able to accurately extract the high $S/N$ SPHERE injection, but systematically underestimated the flux in the GPI data.

The best fit of each algorithm performs relatively well at extracting the true spectrum, with typical best-fit reduced $\chi^{2}/n_{\rm data}$ values approaching 1.
Depending on the injected position angle and separation, the $\chi^{2}$ for the same algorithm at the same separation can vary by a factor of $\sim$10, with typical standard deviations of $\chi^{2}$ on the order of 10$^{1}$--10$^{2}$ depending on the dataset and algorithm.
This variation in extraction suggests that injection recovery tests to measure and correct for algorithm throughput, such as detailed in \cite{greenbaum_gpi_2018}, may introduce additional biases depending on the precise positioning of the injected companions.
In our reproduction of this method, we find that it does not provide better $\chi^{2}$ values, and can introduce spurious wavelength-dependent signals.
This variation is separation-dependent and impacts the extraction less strongly at wider separations, outside the speckle noise regime.

These results point to differences in the approach to data analysis required between detection and characterization efforts.
High numbers of principal components tend to whiten the noise and improve the detection significance, potentially allowing the discovery of fainter companions.
However, this comes at the cost of reduced photometric accuracy, which is critical when attempting to recover the physical atmospheric parameters.
For both speckle subtraction methods, low numbers of principal components produce the most accurate spectral extractions, though the precise number of components will depend on the brightness of the companion in question.

\section{Retrieval Tests}\label{sec:retrievals}
Atmospheric retrievals provide a useful, data-driven tool for exploring the properties of exoplanet atmospheres.
Retrieval results are dependant on the quality of the input data and the assumptions made about both the data and the model.
For our investigation below we have two primary aims:
\begin{enumerate}
    \item Exploring the impact of high-contrast image processing on the inferred atmospheric parameters through retrievals on synthetic data.
    \item Characterising how correlated noise influences the fits of the synthetic data. With a known ground truth, we can explore how the use of the covariance matrix can help mitigate the impact of systematic effects introduced by the data processing.
\end{enumerate}
To this end, we use a representative selection of optimised spectral extractions as described in Section \ref{sec:Injections}.
We choose to use the model injected at 600 mas, and positioned 150\textdegree\ rotated from HR 8799 e, combining both the SPHERE YJH and GPI H-band datasets.
The best extracted spectrum as measured by the relative discrepancy were chosen as the baseline inputs to the retrievals.
This represents a realistic, though challenging spectrum on which to perform atmospheric retrievals.
For validation we also explored a set of retrievals on different locations and choice of extraction, finding that while the precision often varies with the $S/N$ of the extracted spectrum, the overall trends of our results are reproducible.
In contrast to \cite{greco_brandt_2016}, these retrievals will explore the full impact of IFS data processing on the spectra, as opposed to using data synthesised from a parametric estimate of the covariance.
In contrast to their use of a 3-parameter {\tt BT-Settl} model, we use an $\sim$8 parameter forward model in order to understand the cumulative impacts of data post-processing on the inferred atmospheric parameters in the context of high-dimension on-the-fly retrievals. 
Such retrievals are highly flexible, and are more likely to try to fit spurious data features than more physically motivated fits from self-consistent grids.

\subsubsection*{Atmospheric Model}\label{sec:pRT}
The models we use in our atmospheric retrieval setup are computed using \ptrad \citep{molliere_petitradtrans_2019}, a fast, open-source radiative transfer code with which we can calculate the emission spectrum of an atmosphere \footnote{\url{https://petitradtrans.readthedocs.io/}}.
In this framework, the atmosphere of a planet is divided up into pressure bins.
Temperature and chemical structures are calculated and applied to each bin, and radiative transfer using the correlated-k method for the opacities \citep{goody_1989_ck,lacis_1991_ck} is performed to calculate the emission spectrum.
The correlated-k opacities are binned from their native spectral resolving power of 1000 to a user-supplied model resolution using the \verb|exo-k| package \citep{leconte_2021_exok}, improving the computation time of the retrieval.
A wavelength binning of at least twice the data resolution is used for the models, in order that the binned model spectrum will be Nyquist sampled.
This spectral model is convolved with a Gaussian kernel with the width of the instrumental spectra resolution and then binned to the wavelength grid of the input data for the retrieval using the {\tt rebin_giv_width} function.
At the spectral resolutions considered in this work, the effects of the convolution and binning on the spectrum can dominate the spectral shape over data processing effects. 
For this reason we ensured that we use the same convolution and binning procedure during the spectral injections as during the retrieval. 
However, future work should investigate incorporating better instrumental models and wavelength dependent kernels into retrieval frameworks.

Our baseline model uses a Guillot temperature profile \citep{guillot_radiative_2010} and freely retrieved chemical abundances. 
This profile is a simple analytical model, constructed to estimate the thermal structure of irradiated planets:
\begin{multline}
    \label{eqn:modifguillot}
      T_{\rm Guillot} = \frac{3T_{\rm int}^{4}}{4}\left(\frac{2}{3} + \tau\right) \\
            +\frac{3T_{\rm irr}^4}{4}\left(\frac{2}{3} + \frac{1}{\gamma\sqrt{3}} + 
            \left(\frac{\gamma}{\sqrt{3}} - \frac{1}{\gamma\sqrt{3}}\right)e^{-\gamma\tau\sqrt{3}}\right),
\end{multline}
where $T_{\rm irr} = \sqrt{2}T_{\rm equ}$ and $\tau = P\times\kappa_{\rm IR}/g$.
$T_{\rm equ}$ is the standard equilibrium temperature of an irradiated body, $g$ is the surface gravity.
$P$ is the atmospheric pressure, divided up into a total of 80 log-spaced layers from 10$^{3}$ bar to 10$^{-6}$ bar.
The remainder of the parameters are as in \cite{guillot_radiative_2010}: $T_{\rm int}$ is the intrinsic internal temperature of the planet, $\kappa_{\rm IR}$ is the mean infrared opacity, and $\gamma$ is the ratio between the optical and infrared opacities.
All of these parameters are freely retrieved, rather than being derived from the opacities in each atmospheric layer.

This provides a simple but flexible model for the P-T profile, and is the model used to generate the injected spectrum. By setting the irradiation temperature to low values the Guillot model can reproduce the general shape of typical directly imaged planet temperature profiles. Setting the equilibrium temperature $T_{\rm equ}$ to zero provides the limiting case of the Eddington profile \citep{eddington1930}. Together with the planet radius, log $g$, and the chemical abundances this model uses a total of 8 parameters. The only included sources of line opacities are \h2o from the ExoMol data base \citep{chubb_exomolop_2020,tennyson_exomol_2012} and CO from \verb|HITEMP| \citep{rothman_hitemp_2010}.
Both Rayleigh scattering in an H$_{2}$ and He dominated atmosphere and collisionally induced absorption between H$_{2}$-H$_{2}$ and H$_{2}$-He are included as continuum opacity sources. 
The priors used for all parameters in the retrieval are presented in table \ref{tab:RetrievalSetup}.
\begin{table}[t]
    \centering
    \begin{tabular}{lll}
    \toprule
        \textbf{Parameter} & \textbf{Prior} & \textbf{Input}\\
    \midrule
    \multicolumn{3}{c}{Guillot, Free Chemistry}\\
        \midrule
        $\log g$ & $\mathcal{U}\left(2.0, 5.5\right)$ & 4.0\\
        R$_{\rm pl}$ & $\mathcal{N}\left(1.0 \ \mathrm{R}_{\mathrm{jup}}, 0.2 \ \mathrm{R}_{\mathrm{jup}}\right)$ & 1.0 R$_{\mathrm{jup}}$\\[2pt]
        T$_{\rm int}$ & $\mathcal{U}\left(300 {\rm  \ K}, 2000 {\rm  \ K}\right)$ & 750 K\\
        T$_{\rm equ}$ & $\mathcal{U}\left(0 {\rm  \ K}, 300 {\rm  \ K}\right)$ & 100 K\\
        $\gamma$ & $\mathcal{N}\left(1, 0.2\right)$ & 0.5\\
        $\log\kappa_{\rm IR}$ & $\mathcal{U}\left(-3.0, 1.0\right)$ & -1.0\\
        log X$_{\rm H_2O}$ & $\mathcal{U}\left(-7.0, 0.0\right)$ & -1.5\\
        log X$_{\rm CO}$ & $\mathcal{U}\left(-7.0, 0.0\right)$ & -2.0\\
    \bottomrule
    \end{tabular}
    \caption{Priors for retrieval setup. $\mathcal{U}(a,b)$ denotes uniform priors with bounds $a$ and $b$, and $\mathcal{N}\left(\mu, \sigma\right)$ denotes a normal distribution centered at a mean $\mu$ with standard deviation $\sigma$. The final column indicates the true values of the spectrum injected into the IFS cubes.}
    \label{tab:RetrievalSetup}
\end{table}

\subsubsection*{Retrieval Setup}\label{sec:retsetup}
\verb|pyMultiNest| is used to generate samples and determine both the posterior parameter distributions and the Bayesian evidence of the retrieval \citep{buchner_x-ray_2014}.
This is a Python wrapper for the \verb|MultiNest| sampler and likelihood integration method of \cite{feroz_multimodal_2008}.
For all of the retrievals we use 4000 live points to thoroughly explore the parameter space, and a sampling efficiency of 0.8, as recommended in the \verb|pyMultiNest| documentation for parameter estimation.
We compute negative log likelihood, the value of which is minimized in order to find the best-fit set of parameters.
Across many samples, this provides a measurement of the posterior probability distribution of model parameters given the data.
Under the assumption of Gaussian distributed errors, the log likelihood function takes the form of a simple $\chi^{2}$ likelihood distribution.
Using the covariance matrix $\mathbf{C}$ of the data from Section \ref{sec:covar} with elements $\mathbf{C}_{ij}$, we compute the log likelihood function $\log\mathcal{L}$, which is the log-probability of measuring the observed spectrum $\mathbf{S}$ given a forward model $\mathbf{F}$. 
A normalization term is included which allows for a varying covariance matrix or uncertainty for each dataset and penalizes samples with higher uncertainties. 
Thus our likelihood function is computed as:
\begin{equation}\label{eqn:loglike}
    -2\log\mathcal{L} = \left(\mathbf{S}-\mathbf{F}\right)^{T}\mathbf{C}^{-1}\left(\mathbf{S}-\mathbf{F}\right) + \log\left(2\pi\det\left(\mathbf{C}\right)\right).
\end{equation}

\begin{table}
    \centering
    \begin{tabular}{lcccc}
    \toprule
        \textbf{Dataset} & $\boldsymbol{\chi}^{\boldsymbol{2}}/\boldsymbol{ n}_{\textrm{ \textbf{data}}}$ & $\boldsymbol{\log}\mathcal{L}$ & \textbf{BIC} & $\bm{d_{m}}$\\
     \midrule
        \verb|KLIP|, $\mathbf{C}$ & 0.69 & 2692 & -5289 & 2.35 \\
        \verb|KLIP|, diag($\mathbf{C}$) & 0.41 & 2656  & -5218 &  3.29 \\
        \verb|PynPoint|, $\mathbf{C}$ & 1.72 & 2646  & -5220 & 5.63\\
        \verb|PynPoint|, diag($\mathbf{C}$) & 1.43 & 2641  & -5208 & 9.60\\
        \verb|Andromeda|, diag($\mathbf{C}$) & 0.39 & 2650  & -5266 & 4.62\\
        Gaussian, diag($\mathbf{C}$) & 1.06 & 2651  & -5267 &  2.21  \\
        Noise Free, diag($\mathbf{C}$) & 0.06 & 2686  & -5336 & 1.69   \\
    \bottomrule
    \end{tabular}
    \caption{Summary of retrievals run on synthetic data. We compare best fit reduced $\chi^{2}/n_{\rm data}$, the (negative) log likelihood which includes the covariance weighting term of equation \ref{eqn:loglike}, the Bayesian Information Criterion of equation \ref{eqn:BIC} and the Mahalanobis $d_{M}$ from equation \ref{eqn:postdist}. Retrievals were performed on data processed with each algorithm, both using the full and diagonal only terms of the covariance matrix. Toy models using univariate Gaussian scatter about the input and no scatter are also included, with the uncertainties defined as diag($\mathbf{C}$) from the {\tt KLIP} data.}
    \label{tab:SimRetrievals}
\end{table}

\begin{figure*}[t!]
    \centering
    \includegraphics[width=0.48\linewidth]{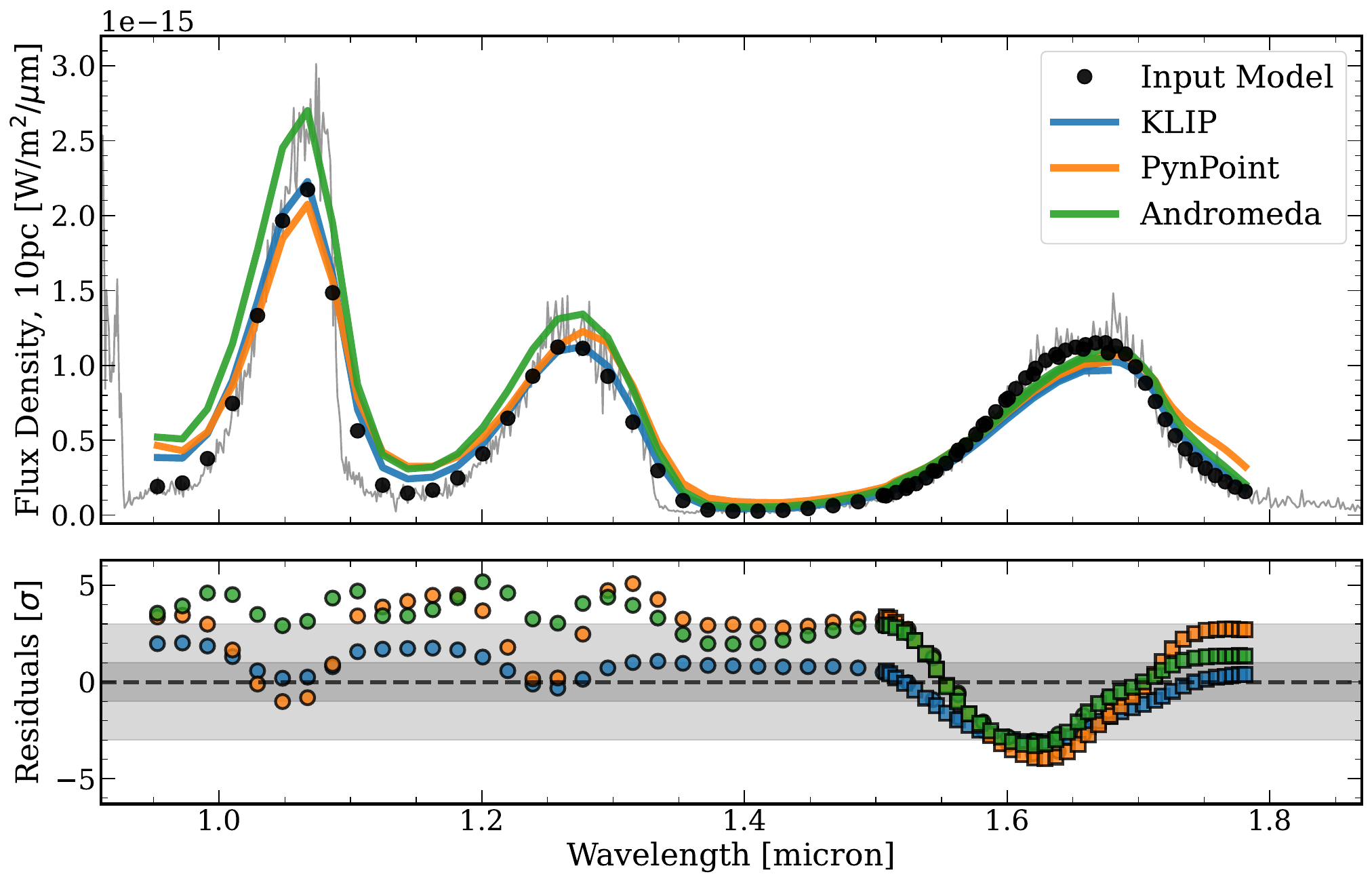}
    \includegraphics[width=0.48\linewidth]{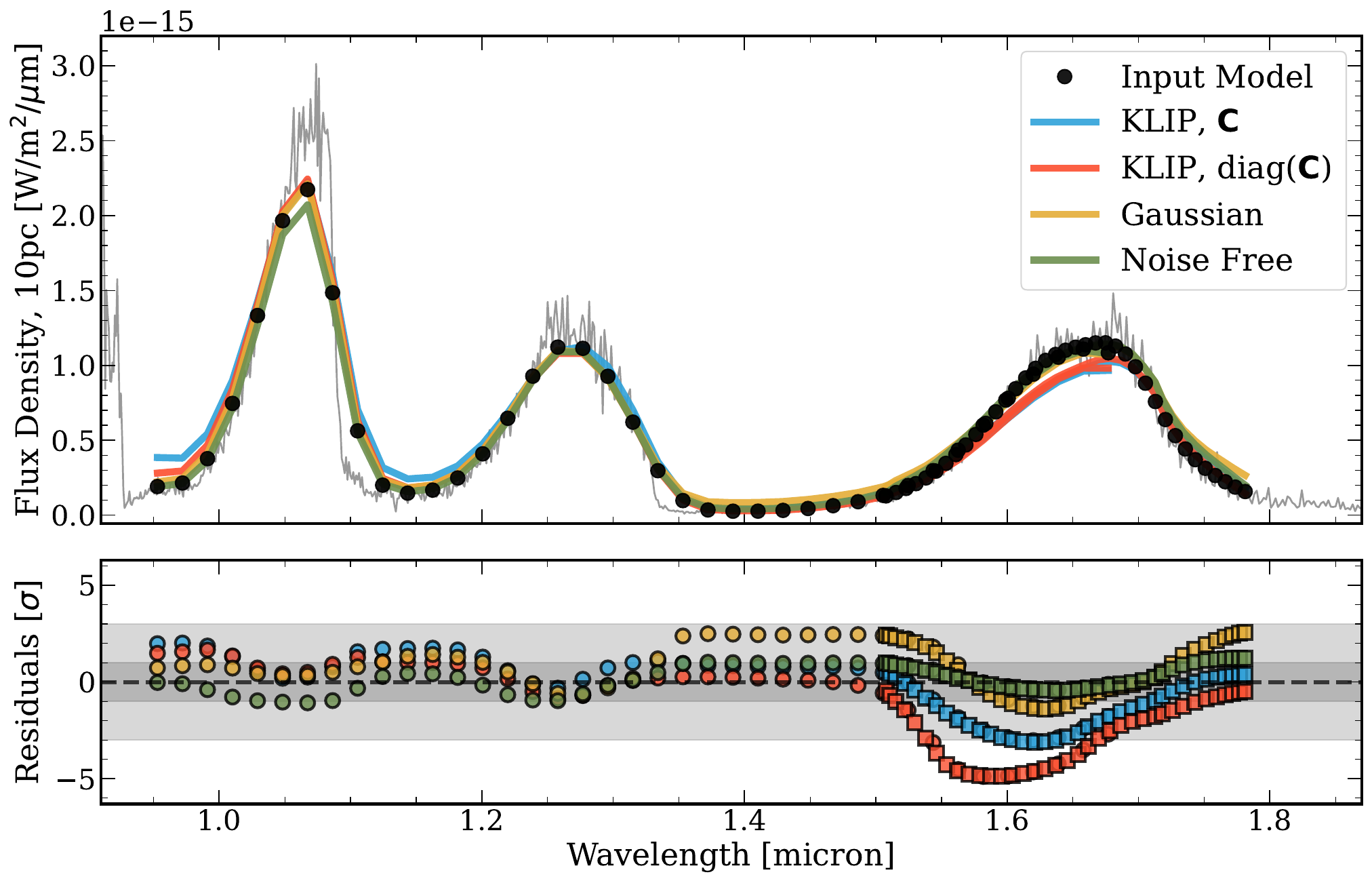}\\   
    \caption{\textbf{Left:} The best fit from the retrievals for each {\tt KLIP} (blue), {\tt PynPoint} (orange) and {\tt ANDROMEDA} (green).
    Each of these fits is compared to the ground truth input model, in order to determine how well the retrieval can account for the systematics introduced through the data analysis. The residuals are calculated by sampling the posterior distributions to generate spectra, and taking the standard deviation at each wavelength. \textbf{Right:} The best fits of the {\tt KLIP} retrievals with (blue) and without (yellow) including the covariance matrix, as well as the retrieval on the Gaussian noise (red) and Noise-free datasets (green). These are again compared to the true input spectrum.}
    \label{fig:KLIPInjSpecs}
\end{figure*}

\subsubsection*{Goodness-of-fit Metrics}\label{sec:retmetrics}
In a retrieval on real exoplanet data without a ground truth value to compare to, we must turn to different metrics in order to determine which retrieval best describes the underlying spectrum.
Table \ref{tab:SimRetrievals} lists $\chi^2$ values of each best fit, as well as the minimum negative log likelihood as computed in Equation \ref{eqn:loglike} and the Bayesian Information Criterion (BIC, \cite{Wit2012_BIC}).
Typically when performing model comparison in a Bayesian framework, we would turn to the Bayes factor in order to reject the null hypothesis.
However, when comparing the impact of different data reductions on the retrieval outcomes, the Bayes factor as computed through the nested sampling evidence estimate is insufficient, as not all of the free parameters are included in the sampling process or in the prior volume, namely the those related to the post-processing algorithms. 
This would bias the evidence estimate, which depends on the choice of priors and thus the overall prior volume.
A full treatment would require marginalizing over these algorithmic parameters, and computing a forward model of the planet signal in the IFS data.
At the present time, such a joint approach is computationally infeasible. 
\cite{ruffio_radial_2019} and \cite{wilcomb2020} demonstrate that this is possible given a linear model of the starlight, the planet signal and the residuals, which can be optimized and analytically marginalised over to determine posterior distributions.
However, this approach loses information on the continuum shape of the spectrum, and relies on moderate-to-high spectral resolution to infer physical quantities. 
The atmospheric model is also not computed on the fly, and instead relies on a precomputed grid, limiting the parameter space available for exploration.

Therefore, instead of the Bayes factor, we rely on the BIC as a summary statistic:
\begin{equation}\label{eqn:BIC}
    \textrm{BIC} = k\log n - 2\log\mathcal{L_{\rm max}}
\end{equation}
for $k$ free parameters and $n$ data points.
This formulation allows us to account for the free parameters of the atmospheric model, as well as the parameters of the data processing, where we add one parameter for each principal component used during PSF subtraction.
Unlike the Bayes factor, the BIC is only a heuristic for model comparison, and differences in the BIC cannot be treated as a metric for statistical significance. 
Nevertheless, models with a lower BIC can be considered more strongly favoured.
As the BIC depends on the likelihood, this also means we cannot directly compare retrievals which include or neglect the off-diagonal terms of the covariance matrix. 
Bayes factors and the BIC estimate whether a certain forward model is favored when compared to another one, whereas turning covariance on or off corresponds to changing the functional form of the likelihood function. 
It is, therefore, not a question of forward model selection.
Thus no single summary statistic can determine the overall goodness-of-fit of the retrieval. 

The $\chi^{2}/n_{\rm data}$ statistic is useful for understanding the impact of varying the covariance and quantifying the similarity of the model to the spectrum, while the BIC is useful for heuristically evaluating the goodness-of-fit, accounting for possible over-fitting from the addition of extra parameters.
In general however, we cannot directly compare the likelihood or the BIC when comparing the cases including the covariance to those using only the diagonal of the matrix with the usual motivation of model selection. 
Adding or neglecting the covariance does not correspond to a different forward model choice, instead it is equivalent to using a correct or incorrect functional form of the likelihood function. 
Therefore, assuming that the covariance is correctly measured, it is \textit{always} better to include the full matrix in the likelihood in order to make statistically robust statements about the the data. 
Thus even though the reduced $\chi^{2}$ of the covariance case may be larger than that with only the diagonal, it still provides a more honest analysis of the data. 
It is also not surprising if the $\chi^{2}$ increases if the covariance is added, since a Gaussian distribution defined by a covariance matrix with non-zero off-diagonal elements will always have a higher information content \citep[e.g.,][]{rodgers2000}. 
As discussed in Section \ref{sec:covimpact}, including the covariance may either increase or decrease the width of the parameter posteriors.

For comparing retrieval results that include or neglect the covariance matrix we make use of the Mahalanobis distance $d_{M}$ \citep{Mahalanobis1936}, to quantify the absolute distance between the posterior probability distributions $P(\theta\,|\,\vec{x})$ with means $\vec{\mu}$ and covariance $S$, and the true parameter values $\vec{\hat{\theta}}$:
\begin{equation}\label{eqn:postdist}
{\displaystyle d_{M}(\vec{\hat{\theta}},P(\theta\,|\,\vec{x}))={\sqrt {(\vec{\hat{\theta}}-{\vec {\mu }})^{\mathsf {T}}S^{-1}(\vec{\hat{\theta}}-{\vec {\mu }})}}.}
\end{equation}
This provides a metric for the overall accuracy of the retrieval when the true input parameters are known.

Finally, for this work we will use the median parameter values and associated spectra as our point of comparison as opposed to the maximum likelihood fit.
We find that although the spectrum generated by the median parameter values is a worse fit (by definition) than the best fit spectrum, the median parameters are a more accurate measurement of the input parameters.

\subsection{Outline of Retrievals}\label{sec:injrets}
We will perform 3 main tests to answer address the central theme of this paper:
\begin{enumerate}
    \item Comparing retrievals on spectra extracted using different post processing algorithms.
    \item Comparing retrievals that either include or ignore the covariance matrix in the likelihood function.
    \item Testing if a lack of correlation information can be accounted for using additional ad-hoc data-processing parameters in the retrieval.
\end{enumerate}

Our primary retrieval results are summarized in Table \ref{tab:SimRetrievals}.
In Section \ref{sec:retalgs} we compare the cases of data that has been processed with {\tt KLIP}, {\tt PynPoint} and {\tt ANDROMEDA}, both with and without the use of the covariance matrix from Section \ref{sec:covar}.
As a benchmark, we also include a retrieval using the nominal input spectrum, perturbed with draws from a Gaussian distribution, where the covariance is given by the diagonal of the {\tt KLIP} covariance matrix.
This represents how the data would appear without systematics from HCI data processing and without the correlations introduced by the instrument optics.
We also include  a retrieval using the same uncertainties as in the Gaussian case, but without scatter about the input spectrum to validate our retrieval method and choice of goodness-of-fit metrics.
For the sake of brevity, we will refer to these as the ``Gaussian'' and ``noise-free'' cases respectively.
We explore the impact of incorporating the covariance matrix in the retrieval framework in Section \ref{sec:retcov}, using the {\tt KLIP}, Gaussian and noise-free cases.
Section \ref{sec:retscales} explores whether we can account for ignorance of the covariance in the data by introducing scaling factors and offsets in the retrieval.

\begin{figure*}
    \centering
    \begin{subfigure}{0.93\textwidth}
    \centering
    \includegraphics[width=0.9\linewidth]{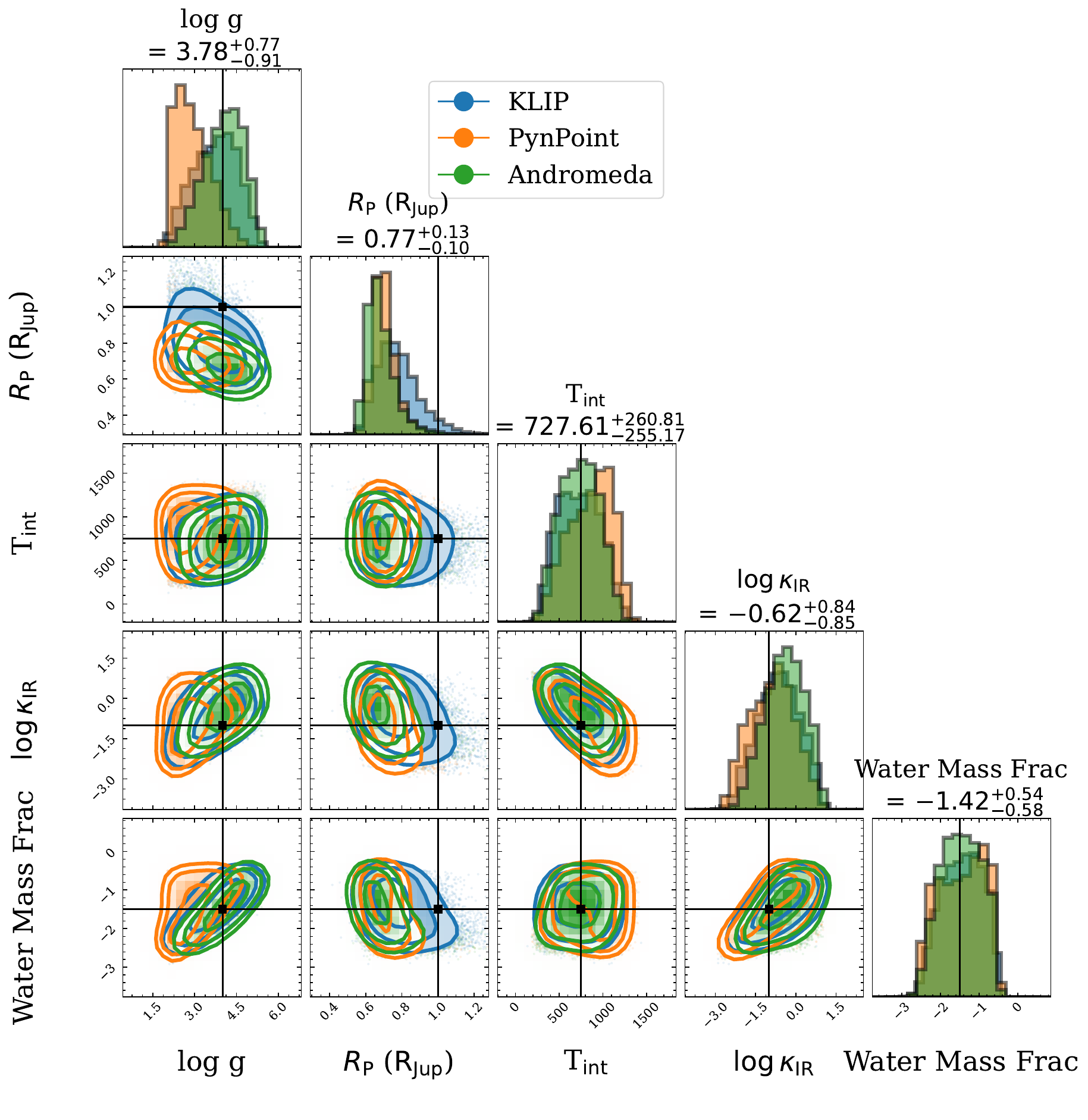}
    \caption{Posterior distributions for retrievals from each data processing algorithm. Contours are plotted for 2D Gaussian 1,2 and 3$\sigma$ levels, corresponding to 36\%, 86\% and 99\% confidence intervals, and the ground truth value is marked in black. The text labels correspond to the {\tt KLIP} retrieval.}
    \label{fig:AlgCompCornerFull}
    \end{subfigure}

    \begin{subfigure}{0.47\linewidth}
    \vspace{-0.2cm}
    \centering
    \includegraphics[width=1\linewidth]{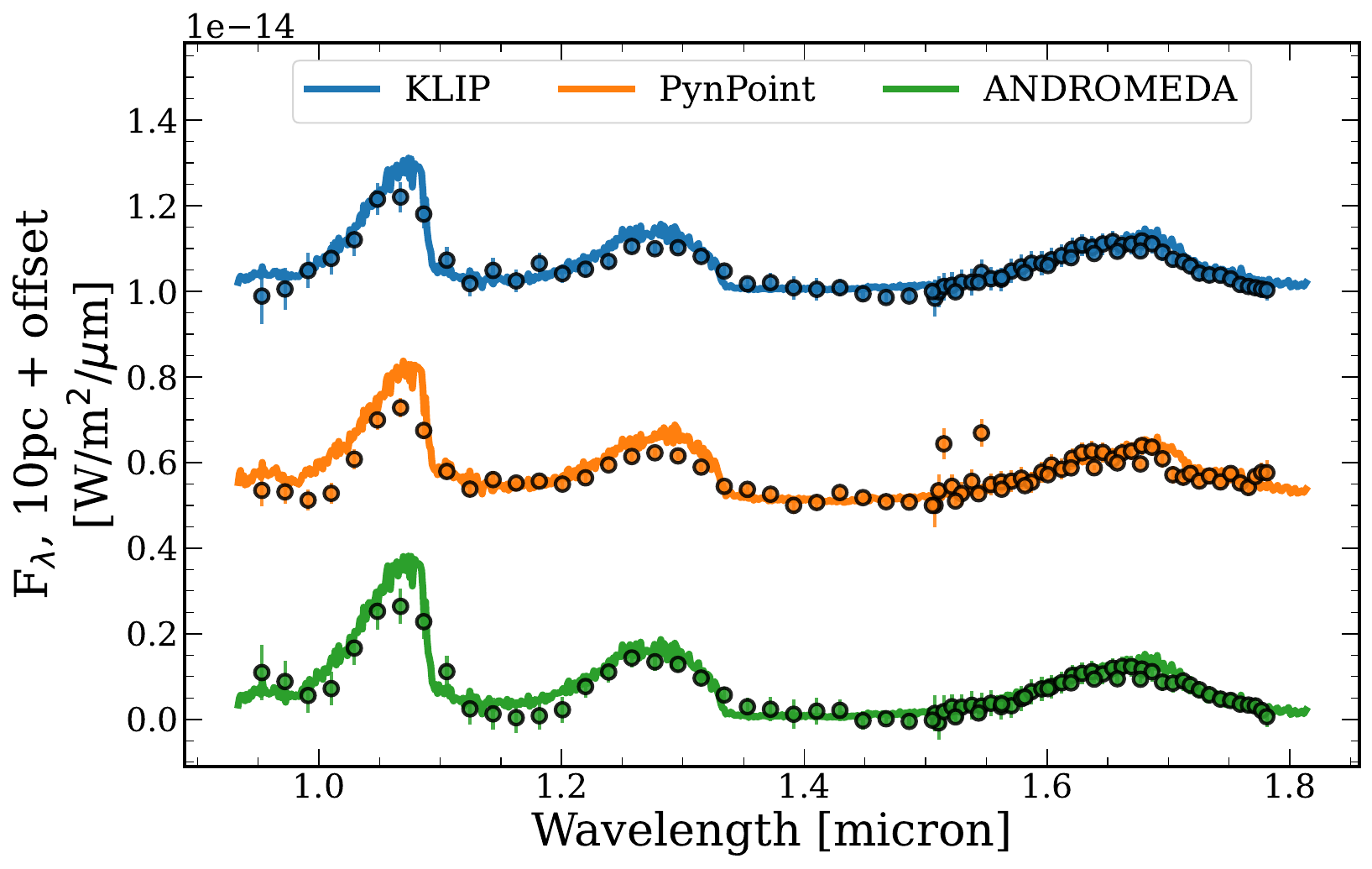}
    \caption{Best-fit spectrum from retrievals on each algorithm, with the covariance.}
    \label{fig:KLIPCovBestFitRetrieval}
    \end{subfigure}
    \begin{subfigure}{0.47\linewidth}\centering
    \includegraphics[width=1\linewidth]{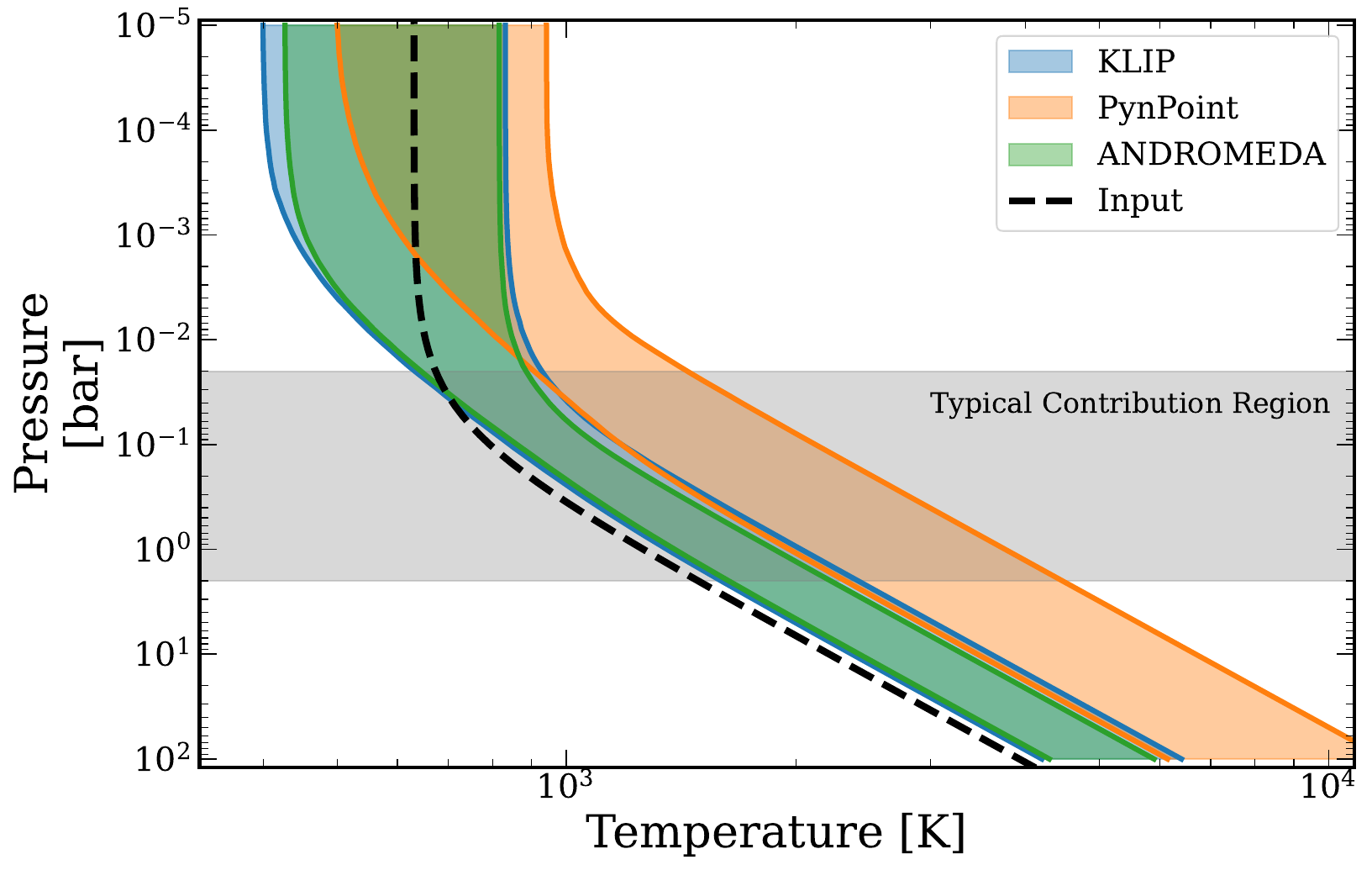}
    \caption{P-T Profile for {\tt KLIP}  retrieval with covariance. The shaded region indicates the 68\% confidence region for the retrieved profile.}
    \label{fig:KLIPCovPT}
    \end{subfigure}
    \caption{Results of retrievals comparing each of the three data processing algorithms. The corner plot given in \subref{fig:AlgCompCornerFull} shows the posterior probabilities of the model parameters. \subref{fig:KLIPCovBestFitRetrieval} shows the median fit retrieval, and \subref{fig:KLIPCovPT} is the associated pressure-temperature profile.}\label{fig:AlgRetrievals}
\end{figure*}

\subsubsection{Validation}
To verify the validity of our results, we also ran a series of validation retrievals to test the sensitivity of our results to the choice of datasets, priors, and models. 
We ran retrievals on each dataset independently, as well as with broad and tight priors.
Neither dataset was able to retrieve the parameters as precisely as the combined retrievals.
The posterior distributions and fits were insensitive to our choice of priors.
We ran additional retrievals using a spline temperature profile, as a proxy for our model not truly matching the underlying data. 
With 5 spline nodes, we were able to retrieve the $\log g$, $R_{\rm pl}$, $T_{\rm int}$ and the water mass fraction to the same precision and accuracy as using the Guillot profile used to generate the data, thus concluding that the retrievals are flexible enough to account for some degree of imperfect model assumptions.

\subsection{How does algorithm selection impact retrieval results?}\label{sec:retalgs}
The largest variation in the extracted spectrum is due to the choice of post-processing algorithm, so our first aim is to explore how these differences in the data lead to differences in the inferred parameters.
In Figure \ref{fig:AlgRetrievals}, we compare the best fit results from each of the retrievals run on data processed using \verb|KLIP|, \verb|PynPoint| and \verb|ANDROMEDA| to the ground truth spectrum injected into the IFS data.
All three processing tools provide reasonable fits to the input spectrum, and share trends in the shape of their residuals, though {\tt KLIP} provides the overall best reproduction of the input spectrum.
The retrieved spectra tend to fit the input better at higher flux values, where the $S/N$ is greater.
Figure \ref{fig:AlgCompCornerFull} shows the posterior distributions of most parameters; regardless of the retrievals setup the absolute uncertainties on all of the retrieved parameters are large. 
This highlights the importance having high $S/N$ inputs to obtain precise constraints, as well as broad wavelength coverage to have sensitivity to a wide range of parameters.
Not included in the plot are T$_{\rm equ}$, $\gamma$ and the CO mass fraction, none of which are constrained in any retrieval. 
For widely separated planets, T$_{\rm equ}$ is small and has little impact on the shape of the pressure temperature profile.
As such, the Guillot profile is effectively reduced to the Eddington term, which does not depend on $\gamma$.
Finally, there are no strong CO features present in the wavelength range considered in the injected planets. 
Thus we do not expect any of these parameters to impact the spectrum enough to be constrained by this retrieval.
We neglect these unconstrained parameters when calculating the distance $d_{M}$.

{\tt KLIP} performs the best of the three algorithms; accurately fitting the spectrum and retrieving the input parameters, measured from the $\chi^{2}/n_{\rm data}$ and $d_{M}$ respectively, as presented in table \ref{tab:SimRetrievals}.
{\tt PynPoint} performs somewhat worse again, though the GPI data suffers from two outlying data points, and the measured uncertainties are generally smaller than for the {\tt KLIP} or {\tt ANDROMEDA} extractions. 
The physical interpretation of the {\tt PynPoint} is significantly different than that of {\tt KLIP} or {\tt ANDROMEDA}: the inferred mass from the median $\log g$ and planet radius is more than a factor of 10 smaller when using the {\tt PynPoint parameters}.
This highlights the need for feedback between modelling and data analysis, as well as for comparison both between different data analyses and different models.
This strongly impacts the measurement of $\log g$, which is sensitive to the shape of the H-band.
Finally {\tt ANDROMEDA} fails to reprocuce the input spectrum well, but recovers the input parameters more accurately than {\tt PynPoint}, though confidently excluding the true planet radius.

We find that for all of the retrieval setups, the median parameter values provide a better estimation of the true input parameters than the single maximum likelihood fit. 
For {\tt KLIP}, we find that the median internal temperature estimate of $724\pm260$ K accurately, if imprecisely measure the true value of $750$ K. 
However, the best fit value of 498 K is strongly biased from the true value, as are the remaining parameters.
We therefore continue using only the median parameter estimates, rather than the maximum likelihood fits.

The goodness-of-fit metrics provide a mechanism to select between the different retrievals.
We find that all of the metrics favour the {\tt KLIP} retrieval, though noting that the reduced $\chi^{2}$ for {\tt ANDROMEDA} is smaller, but does not account for the covariance.
Based on the variation in the BIC and the Mahalabois distance, the effect that the algorithm choice has on the retrieved parameters is significant.
Interpreting the Mahalabois distance as standard deviations from the truth, {\tt ANDROMEDA} and {\tt PynPoint} are 2.3 and 3.3 standard deviations less accurate than {\tt KLIP} respectively.
The trend of the BIC follows that of $d_{M}$, favouring {\tt KLIP}, followed by {\tt Andromeda} and then {\tt PynPoint}.
As the ground truth is not generally known, this reinforces the use of the BIC or a similar metric (such as the Bayes factor) as a robust metric for selecting between models, even when the data is also varied.

None of the retrievals retrieve the true input pressure temperature profile to within 1$\sigma$, as is evident from Figure \ref{fig:KLIPCovPT}. 
However, in the region where the emission contribution is located, the retrieved PT profiles share a similar slope to the true input, at slightly higher temperatures. 
Such discrepencies highlight the importance of broad wavelength coverage in atmospheric retrievals, where the spectrum can probe different pressure, and thus temperature, layers of the atmosphere.

\subsubsection{How important is choosing the optimal number of  principal components?}\label{sec:retpca}
We also compare how the number of principal components used in the data processing impacts the retrieval results.
This effect is more apparent at lower $S/N$ so for this particular case we choose an injection at 400 mas, extracting the spectra with {\tt KLIP}. 
Figure \ref{fig:nPCsKLIPCorner} highlights the impact that the choice of the number of PCs has on the precision of the posterior distributions.
We compared the optimal extraction (6PCs for GPI and 8 for SPHERE) to an extraction using 25 PCs for each dataset.
While both extractions retrieve the input parameters with similar accuracy, the optimised extraction is significantly more precise in its measurement of the planet radius.
\begin{figure}[t!]
    \centering
    \includegraphics[width=\linewidth]{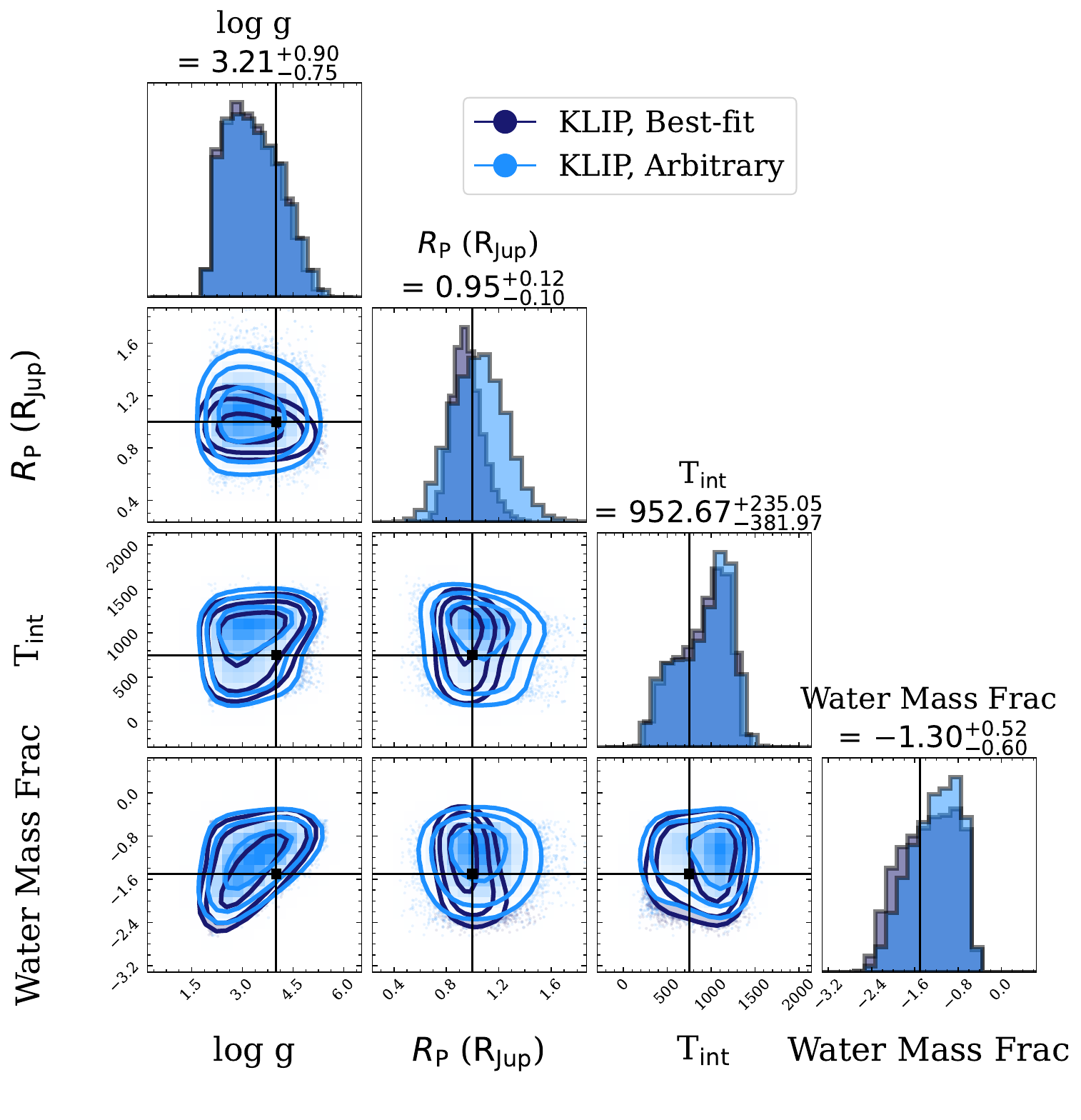}
    \caption{Corner plot comparing the retrieved parameter distributions for two different {\tt KLIP} reductions. In light blue, the input spectrum was optimized using the relative discrepancy metric (Figure \ref{fig:Chi2SeparationPC}), while in dark blue an arbitrary extraction was chosen for each of the SPHERE and GPI datasets, reflecting a non-optimal parameter selection.}
    \label{fig:nPCsKLIPCorner}
\end{figure}

\begin{figure*}
    \centering

    \begin{subfigure}{0.9\linewidth}
    \centering
    \includegraphics[width=0.9\linewidth]{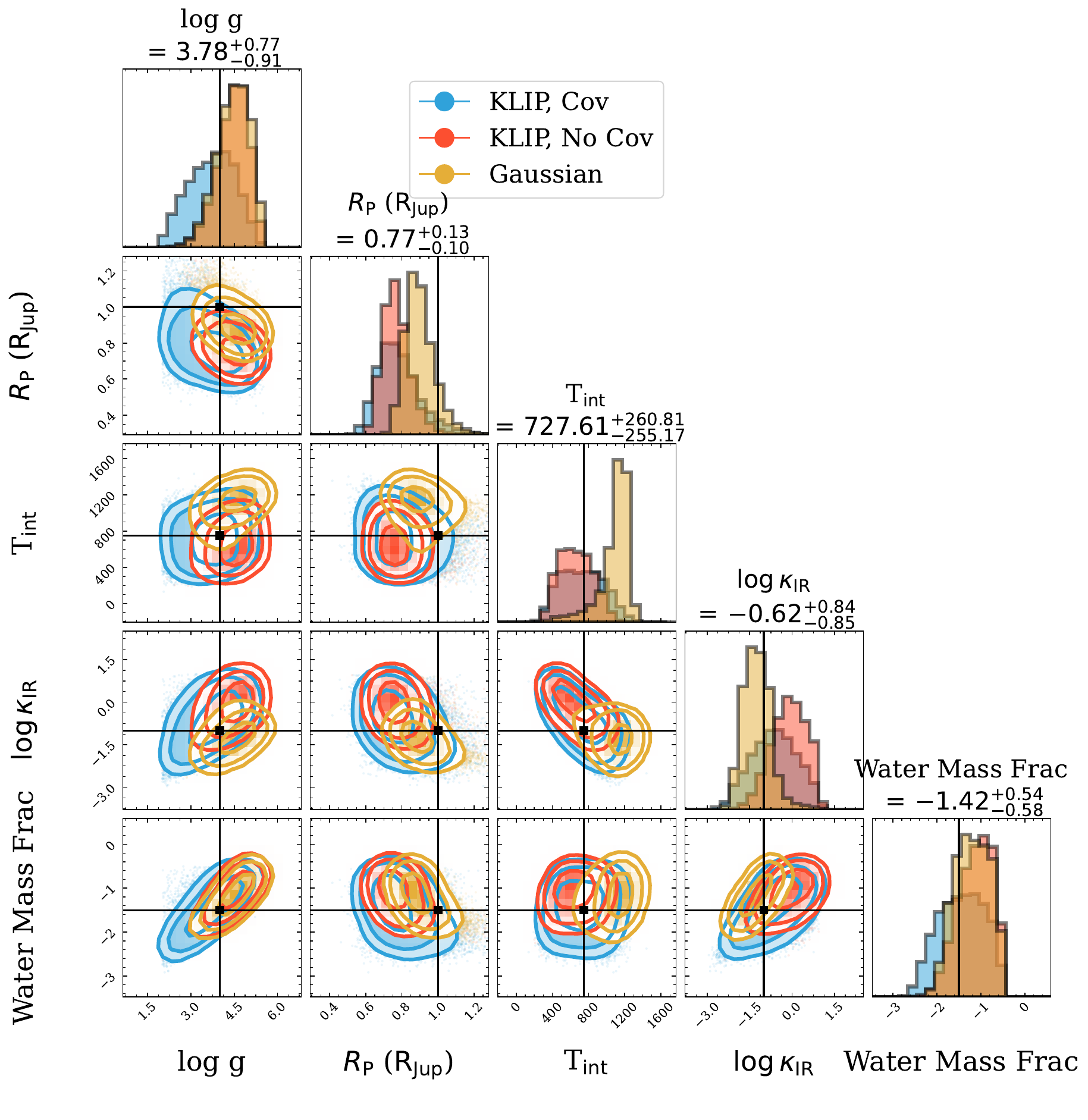}
    \caption{Posterior distributions for retrievals from {\tt KLIP}. Contours are plotted for 2D Gaussian 1,2 and 3$\sigma$ levels, corresponding to 36\%, 86\% and 99\% volume regions, and the ground truth value is marked in black. The text labels correspond to the retrieval including the covariance matrix.}
    \label{fig:CovCornerFull}
    \end{subfigure}
    
    \begin{subfigure}{0.47\linewidth}
    \vspace{-0.25cm}
    \centering
    \includegraphics[width=1\linewidth]{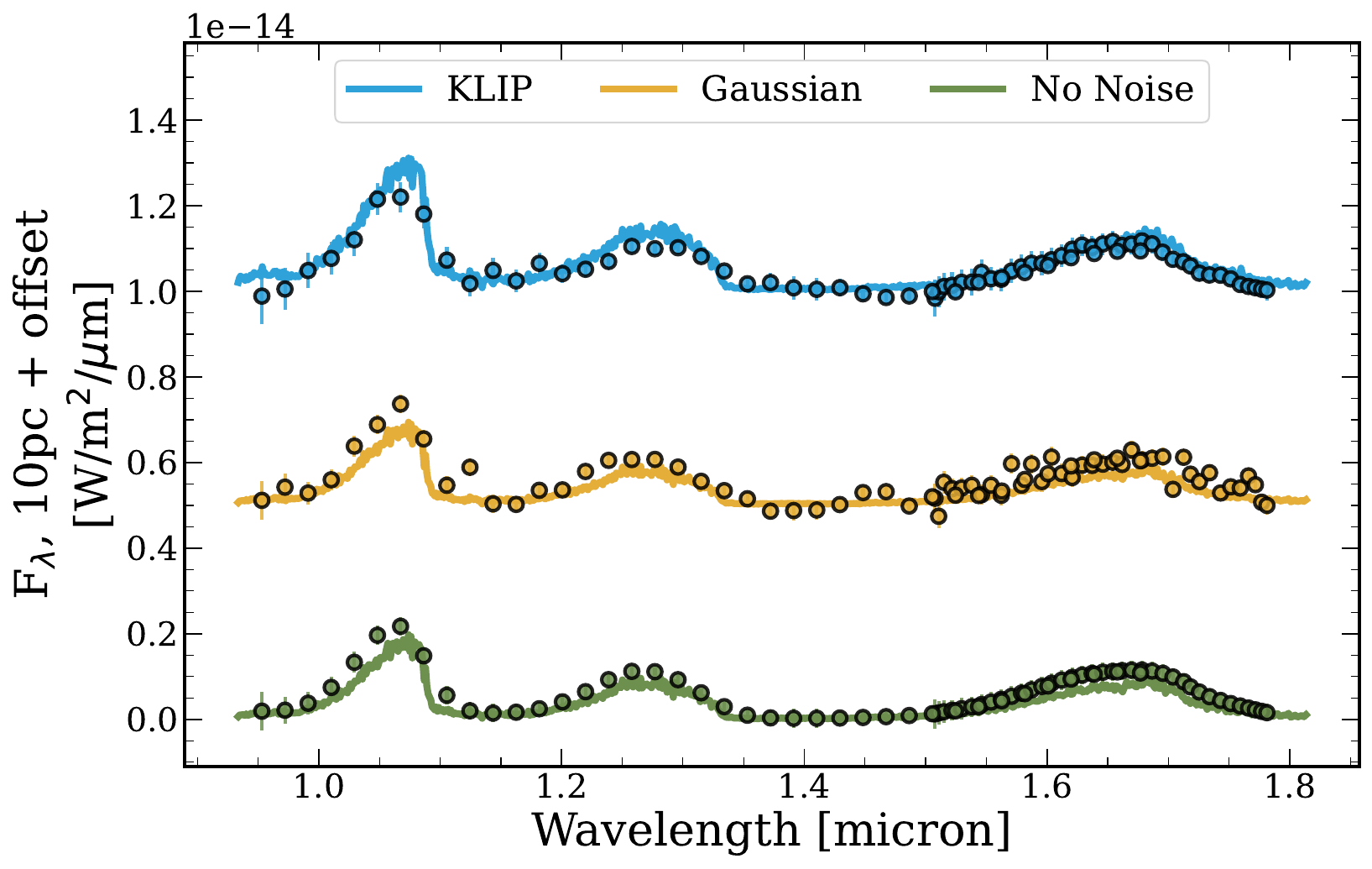}
    \caption{Best-fit spectrum for the {\tt KLIP} retrieval with covariance, as well as for the Gaussian (yellow) and noise free cases (green).}
    \label{fig:CovBestFitRetrieval}
    \end{subfigure}
    \begin{subfigure}{0.47\linewidth}\centering
    \includegraphics[width=1\linewidth]{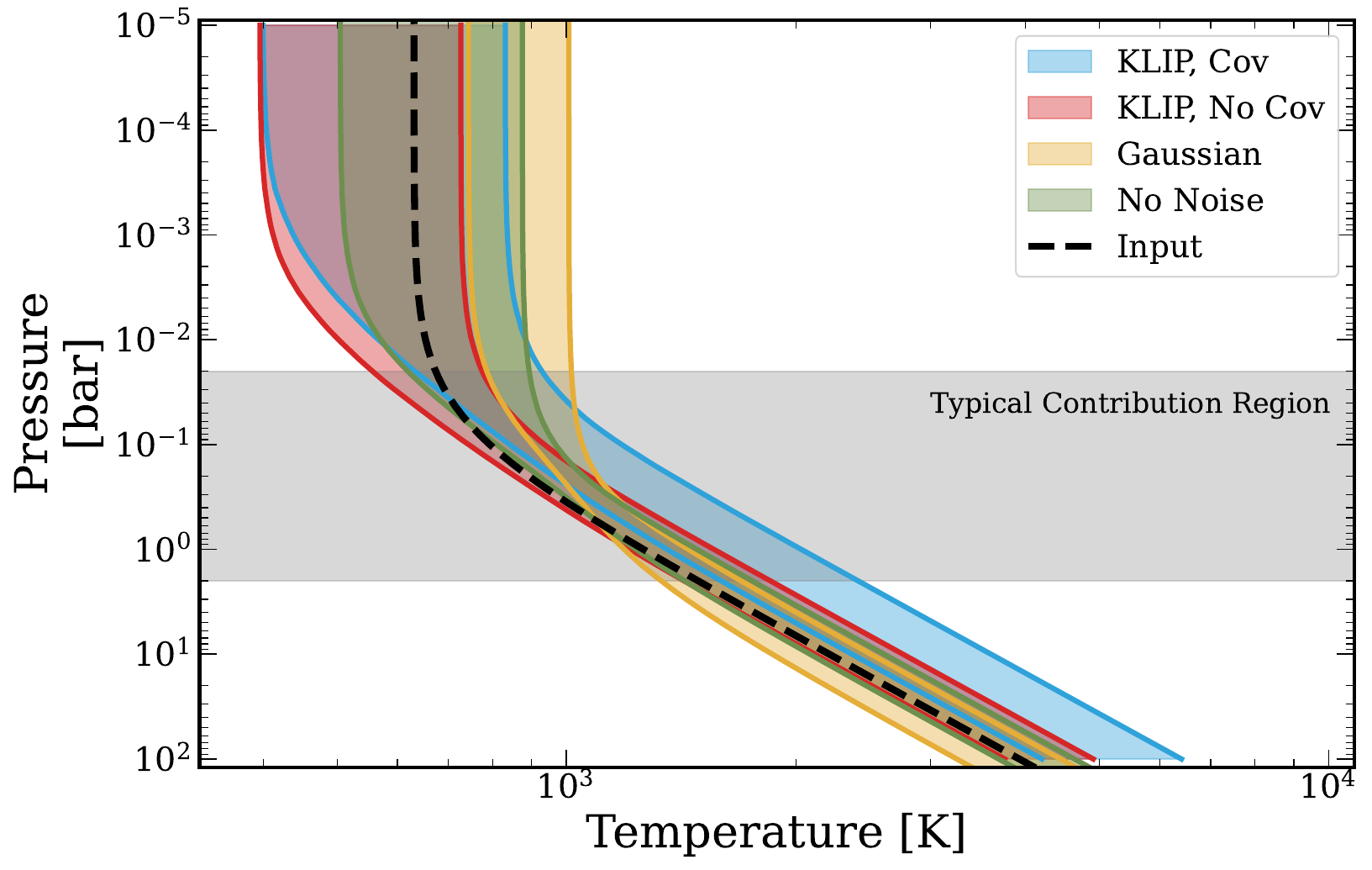}
    \caption{P-T Profile for {\tt KLIP}, Gaussian, and noise-free retrievals. The shaded region indicates the 68\% confidence region for the retrieved profile.}
    \label{fig:CovPT}
    \end{subfigure}
    \caption{Results of retrievals using {\tt KLIP}, comparing each the cases of computing the likelihood using the full covariance matrix (blue), the diagonal elements only (yellow) and using truly Gaussian scattered data (red). The corner plot given in \subref{fig:CovCornerFull} shows the posterior probabilities of the model parameters. \subref{fig:CovBestFitRetrieval} shows the median retrieved spectra, and \subref{fig:CovPT} is the associated pressure-temperature profile.}
    \label{fig:covretallpanels}
\end{figure*}

\begin{figure}
    \centering
    \includegraphics[width=\linewidth]{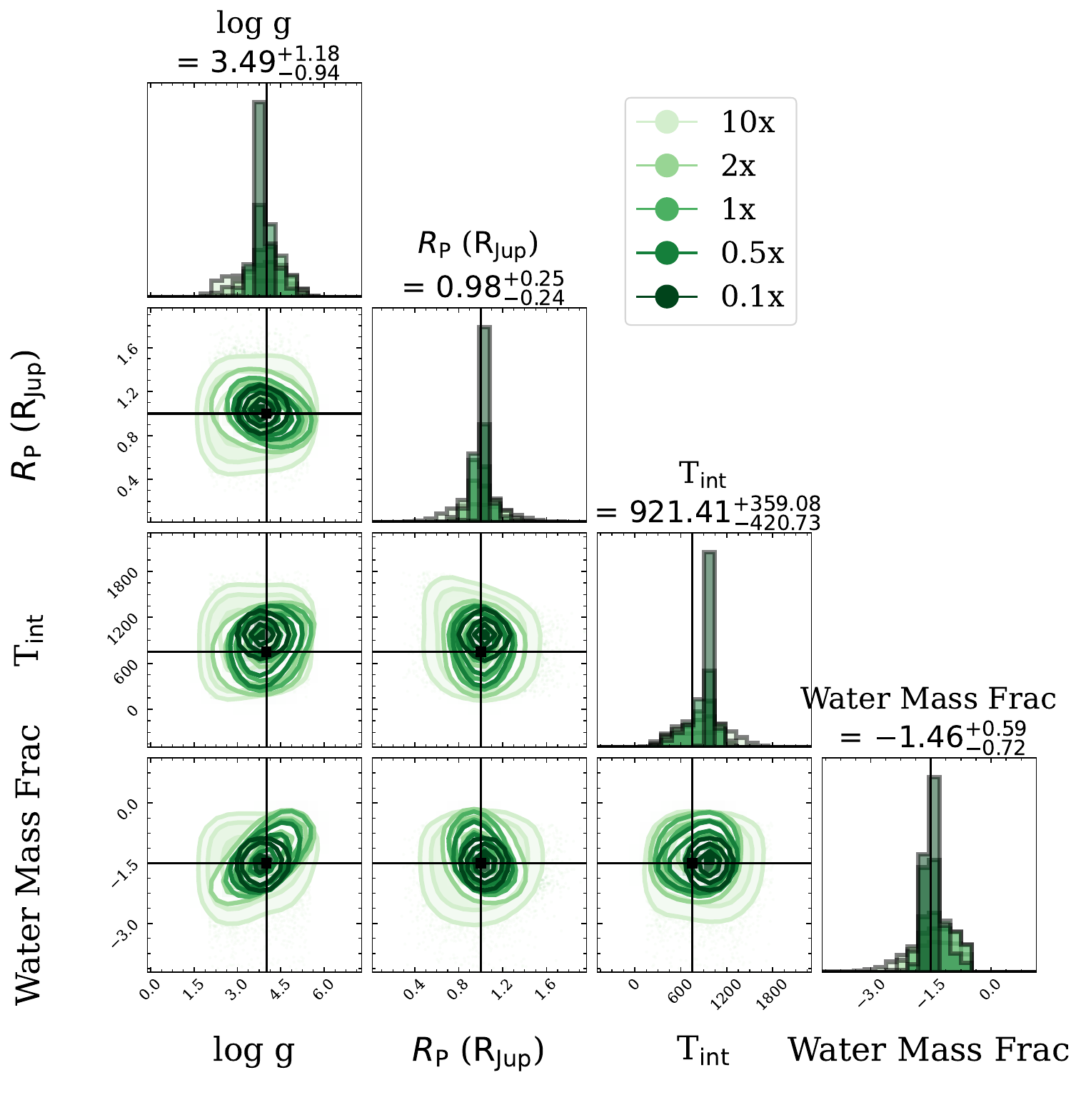}
    \caption{Posterior distributions for retrievals in the `Noise-free' case. The nominal uncertainties are from diag(${\vec{C}}$) of the {\tt KLIP} extraction, and were scaled by factors from 0.1 to 10. The titles list the uncertainties for the 10x case.}
    \label{fig:NoiseFreeCorner}
\end{figure}

\subsection{How does covariance impact retrieval posteriors?}\label{sec:retcov}
While the choice of algorithm produces most of the difference in the spectral shape, we also consider how including the covariance into the log-likelihood calculation of the retrieval impacts the retrieved spectrum and inferred parameters.
To understand this, we compare a \verb|KLIP| retrieval with and without the use of the covariance matrices, and how these results compare to the Gaussian and noise-free cases.

The right panel of Figure \ref{fig:KLIPInjSpecs} shows each of these datasets compared to the true input model, while Figure \ref{fig:covretallpanels} shows the posterior distributions, best fits and PT profiles.
We find that our retrievals reproduce the results of \cite{greco_brandt_2016}: incorporating the covariance matrix improves the accuracy of the retrieval, at the cost of lower precision.
Including the covariance matrix reduced the posterior bias for all parameters.
However, the diag($\vec{C}$) case was able to more accurately retrieve the input PT profile, even though the parameter distributions were more discrepant from the true input parameters.

Quantitatively, including the covariance improved the $d_{M}$ by 1 compared to the diag($\vec{C}$), approaching the Gaussian measurement of 2.21.
As in Section \ref{sec:covimpact}, the $\chi^{2}/n_{\rm data}$ is underestimated.
This is also reflected by the BIC, which favours the inclusion of the covariance matrix over both the diag($\vec{C}$) case and the Gaussian case.
In the {\tt KLIP} extraction of the GPI data, there is rather broad covariance, which is easier to fit, as demonstrated by the toy model of  \ref{sec:covimpact}.

Within the pressure range probed, all of the retrievals measure similar slopes, and correct temperatures to within 2$\sigma$, with the noise-free and diag($\vec{C}$) cases most accurately retrieveing the input profile.
The correlation between atmospheric parameters of $T_{\rm int}$ and $\kappa_{\rm IR}$ in Figure \ref{fig:CovCornerFull} shows the difficulty in inferring atmospheric properties, and explains the inability of the retrievals to perfectly infer the temperature structure.
For the {\tt KLIP} data, the inclusion of the covariance matrix in the log-likelihood improves the accuracy of the constraint on both parameters.

We repeated this experiment using the \verb|PynPoint| extractions.  Consistent with the \verb|KLIP| results, we find that including the covariance improves the accuracy of the retrieval, at a marginal cost to the precision of the retrieved parameters.

\subsubsection{How does measurement precision impact the posterior precision?}
With the nominal uncertainties, the noise-free case can reproduce the input data to within 1$\sigma$ across the entire wavelength range.
However, at the level of precision of these measurements none of the retrievals are able to put strong constraints on any of the measured parameters, though this would be improved with more precise spectroscopic measurements.
Using the noise-free case, we explored how the measurement precision affects the posterior precision by scaling the uncertainties by factors from 0.1 to 10, shown in Figure \ref{fig:NoiseFreeCorner}.

In all cases, the parameters are accurately retrieved, and the posterior precision increases as the uncertainties decrease.
In the nominal case, the mean $S/N$ per channel is 5 in the SPHERE wavelength range and 3 in the GPI-H band.
With this precision and wavelength coverage, even in the optimistic case of no scatter in the data the internal temperature can only be constrained to within $\pm 223$ K. 
This improves to $\pm 26$ K if the $S/N$ is improved by a factor of 10: while low $S/N$ may be sufficient for accurate retrievals, high $S/N$ is required to precisely measure the physical parameters.
This finding complements Figure \ref{fig:nPCsKLIPCorner}, highlighting that the main impact of the principal component optimisation is to improve the precision of the posterior distributions, as the choice of PCs impacts the precision of the spectroscopic measurement.
Our choice of metrics are also validated, as the noise-free case is favoured by every metric (when comparing to diagonal only cases).

\begin{table}
    \centering
    \begin{small}
    \begin{tabular}{lccccc}
    \toprule
        \textbf{Dataset} & $\boldsymbol\chi^{\boldsymbol2}/\boldsymbol n_{\textrm{ \textbf{data}}}$ & $\boldsymbol{\log}\mathcal{L}$ & $\boldsymbol\log_{\mathbf{10}}\boldsymbol{\mathcal{Z}}$ & \textbf{BIC} & $\boldsymbol{d_{M}}$ \\
     \midrule
        \verb|KLIP|, $\mathbf{C}$ & 0.69 & 2692 & 1166 & -5289 & 2.35 \\
        \verb|KLIP|, diag($\mathbf{C}$)  & 0.41 & 2656 & 1149 & -5218 &  3.29 \\
        Offset GPI     & 0.42 & 2657 & 1147 & -5214 & 3.29\\
        Offset SPH     & 0.36 & 2659& 1148 & -5218 & 3.40\\
        Scale GPI      & 0.51 & 2666 & 1152 & -5233 & 4.48\\
        Scale SPH      & 0.56 & 2659 & 1149 & -5218 & 3.93 \\
        Scale GPI Err. & 1.07 & 2684 & 1158 & -5268 & 4.52\\
        Scale SPH Err. & 0.70 & 2660 & 1150 & -5220 & 4.73\\
        Scale Both.    & 1.15 & 2684 & 1158 & -5263 & 4.54\\
        $10^b$ Both.   & 0.43 & 2567 & 1148 & -5210 & 3.41\\
        GPI Only, $\mathbf{C}$& 0.51 & 1320 & 571 & -2596 & 0.62\\
        SPH Only, $\mathbf{C}$& 1.00 & 1373 & 593 & -2698 & 2.75 \\

    \bottomrule
    \end{tabular}
    \end{small}
    \caption{Summary statistics for {\tt KLIP} retrievals, including retrieved parameters to account for systematic biases. $\log_{10}\mathcal{Z}$ is the Bayesian evidence, the difference of which is the Bayes factor between two models. "Scale" indicates a multiplicative factor applied to the specified dataset, while "Offset" indicates an additive term.}
    \label{tab:ScaleFactor}
\end{table}
\subsection{Can we retrieve nuisance parameters to account for systematics?}\label{sec:retscales}
One challenge to the interpretation is the suggestion of overfitting by the reduced $\chi^{2}$ values. 
A $\chi^{2}/n_{\rm data}<1$ suggests overfitting, although it can also be interpreted as overestimated uncertainties, or underestimated correlation (which effectively translates into overestimated uncertainties as well).  
The retrievals both with and without covariance on the \verb|KLIP| dataset both have $\chi^{2}/n_{\rm data}<1$.  
As the number of parameters is much lower than the number of data points and is identical to the ground truth model, this suggests that the uncertainty of the \verb|KLIP| data is overestimated.  
In this section we explore the use of various parameterisations to account for systematics in the data, such as including offsets or scaling factors in the retrievals.  
As all of these comparisons use the same data, and additional parameters are properly included in the prior volume, we can now use the Bayes factor from Table \ref{tab:ScaleFactor} to quantitatively select the best model.
We refer to Table 2 of \cite{benneke_2013_bayes} for our interpretation of the Bayes factor: $\log\Delta\mathcal{Z}_{H_{2},H_{1}}>10$ is strong evidence in favour of model $H_{2}$ over $H_{1}$.

Beginning with offsets, we fix one dataset and allow the other dataset to float, with a uniform prior of $\mathcal{U}(-10^{-14}\textrm{W/m}^{2}/\upmu\textrm{m}, 10^{-14}\textrm{W/m}^{2}/\upmu\textrm{m})$. 
We find that while allowing for offsets may marginally improve the fit to the data ($\chi^{2}_{\rm SPH,\, offset}/n_{\rm data}<\chi^{2}_{\rm KLIP,\, diag(C)}/n_{\rm data}$), it does not improve the accuracy or precision of the posteriors ($d_{M}>d_{M,\,\rm{KLIP,\, diag(C)}}$), and is not favoured by the Bayes factor or BIC. 

Next, we multiply one dataset and its corresponding uncertainties by a scaling factor ($\mathcal{U}(0.5,2.0)$), fixing the remaining dataset.
We find that a scaling factor of 0.73$\pm$0.05 for the GPI dataset is somewhat favoured by the Bayes factor ($\Delta\log_{10}\mathcal{Z}=3$), though the posterior precision and accuracy is somewhat reduced.
Scaling the SPHERE data did not significantly improve the fit, and is not favoured by the Bayes factor.

This result is emphasised when we scale only the uncertainties for each of the datasets.
We ran three retrievals: scaling the uncertainties of each dataset and fixing the other, or scaling both datasets with independent scaling factors simultaneously.
To avoid hitting the prior boundaries, the scaling factor is given a uniform prior of $\mathcal{U}(0.05,2.0)$.
We find that a scaling factor of 0.28$\pm$0.04 is favoured ($\Delta\log_{10}\mathcal{Z}=9$) for the GPI dataset in both retrievals where the uncertainties are allowed to float, while allowing the SPHERE uncertainties to float does not change the fit.
This implies that the either the {\tt KLIP} uncertainties for the GPI dataset are underestimated, or that the correlation is not correctly accounted for, as described in Section \ref{sec:covimpact}.
Allowing the uncertainties to float improves the fit compared to the retrieval using only the diagonal components of the covariance matrix, but is still disfavoured compared to the retrieval using the full matrix.

\cite{line_uniform_2015} introduces a different parameterisation to scale the uncertainties and reduce overfitting.
Using a parameter $b$, the uncertainty on the $i^{th}$ wavelength bin is inflated as:
\begin{equation}\label{eqn:line}
    s_{i}^{2} = \sigma_{i}^{2} + 10^{b},
\end{equation}
where $\sigma_{i}$ is the uncertainty on that bin.
This will only allow for an increase in the size of the error bars, and allows us to account for model uncertainties and missing physics, rolling the additional uncertainty into the marginalized posterior parameter distributions.
The prior range on $b$ is set from -36 to -26: this encompasses the suggested range from \cite{line_uniform_2015} such that $0.01\times\min(\sigma^{2}) < 10^{b} < 100\times\max(\sigma^{2})$.
We retrieve $b$ independently for each dataset used in the retrieval.
Using this formalism, we find that the $b$ parameterised retrieval is disfavoured compared to the \verb|KLIP| retrievals, both with and without the use of the covariance matrix.  
Likewise, the $d_{M}$ measured for this case (3.41) suggests a marginal decrease in accuracy relative to the baseline retrievals. 
This suggests that the extra parameters used to fit the $b$ parameter are not justified to help resolve the problem of overfitting of the spectrum.
As this formalism can only inflate the uncertainties, it is unsurprising that it cannot correct for the overestimated GPI uncertainties as shown by the scaling factor retrievals and the small reduced $\chi^{2}$ values.

We conclude that scaling factors and offsets are inadequate for accounting for systematic offsets in the data due to the data processing and correlated noise.
While allowing the uncertainties to float was marginally favoured by the Bayes factor compared to the diag($\vec{C}$) {\tt KLIP} retrieval, it was still strongly disfavoured compared to the retrieval using the full covariance matrix.
Alternative methods, such as Gaussian Process regression \citep{Wang2021PDS70, xuan2022_cloudybds} may be able to overcome these limitations and allow for the characterisation of systematics in a Bayesian framework. 

\subsection{Limitations}
This work reflects many of the best-practices used in both data analysis and atmospheric retrievals, but remains an optimistic assessment of our ability to infer both accurate and precise physical parameters.
Additional sources of bias are inevitably present in the data, such as the differences in spectra arising from different reduction pipelines as shown in appendix \ref{sec:sphereappendix} and from the process of building the 3D cubes from the 2D detector frames.
Exoplanet data relies heavily on precise photometry, yet the host star which is used as a calibration source is obscured behind the coronagraph during the observations, making temporal monitoring of the PSF challenging.
Finally, we use the same model for both injection and as the basis of the retrieval, ultimately ignoring many key physical processes present in real exoplanets.
Even with these limitations, we remain optimistic about the prospects for retrievals to characterise directly imaged exoplanets, particularly in the era of high precision, broad wavelength spectroscopy as enabled by VLTI/GRAVITY, JWST and the ELTs.

\section{Summary \& Conclusions}\label{sec:discussion}

Based on our comparison of high contrast imaging algorithms, it is clear that systematic variations are a more significant contribution to uncertainty than random errors for directly imaged exoplanet spectra.
Such variations often lead to spectral correlation of the data, and knowledge of the  length scale and strength of said correlation is crucial to accurate interpretation of of the data.
We used the methods of \cite{greco_brandt_2016} to compute covariance matrices for IFS data, and demonstrated that correlations in the data can both increase or decrease the posterior width of model parameters, depending on whether the parameter is sensitive to wavelength scales greater or less than the correlation scale.
Using injection testing, we optimize our choice of algorithm parameters.
We find that using only the $S/N$ as a metric to determine the quality of spectral extraction does not produce optimal extractions.
Instead, data processing parameters should be tuned using injection testing, with careful consideration of what goodness-of-fit metric should be used.
Using the mean relative discrepancy, we optimized the number of principal components used in PSF subtraction in order to optimally extract the companion spectrum.
Of course the number of principal components used in PSF subtraction is not the only source of systematic biases during spectral extraction: the precision of the astrometric solution, choices in processing both the science frames and the unsaturated PSF frames and the details of parameter choice all introduce biases on a similar level to the number of principal components, and must be independently optimized.

Each algorithm considered performed best under different conditions: the contrast, separation, observing conditions and data volume all impact which algorithm produces the optimal extraction. 
Care must be taken as the parameter choices that lead to the the most sensitivity in order to detect companions are often different than the parameters required to robustly extract the spectrum in order to characterise the planets.
During independent comparisons such choices led to statistically significant differences in both the shape and overall flux calibration of the extracted spectra. 
Without a priori knowledge of the spectrum, it is therefore necessary to compare multiple independent measurements in order to determine the underlying spectral shape.

By performing atmospheric retrievals on data processed using different algorithms we show that the variation between different data reductions is larger than the statistical posterior uncertainty.  
Model choice is highly dependant on data quality and quantity, and Bayesian comparisons should be performed to determine whether model complexity is suitable given the data.  
The ideal solution is to fully understand and correct for systematics during the data processing, with broad data coverage and high spectral resolution.  
When comparing models, statistical tools such as the BIC or the Bayes factor should be used with care, and only when the free parameters are fully incorporated into the retrieval process to account for their impact on the prior volume and posterior distribution.
In such a Bayesian framework, we find that the median parameter values are accurate measurements of the true input parameters, but that the maximum likelihood values are often strongly biased.
Even using the median values, the difference in retrieved parameters from different data processing tools is significant, and can lead to dramatically different astrophysical interpretations.
Retrievals should be performed on multiple data reductions to ensure that the retrieved parameters are robust to such variation.

We used the Mahalanobis distance, $d_{M}$ to measure the distance between the true input parameters and the posterior distributions. Using this metric, we found that accounting for the covariance in the likelihood function of a retrieval framework can help mitigate correlations in the data, but not entirely resolve them.
Compared to using only the diagonal terms using the full matrix will reduce the bias in all parameters, at the cost of slightly decreased precision, reproducing the results of \cite{greco_brandt_2016}. 
When the data are correlated including the covariance is necessary to make meaningful statistical statements about models fitting the data. 
In all cases, including the full covariance matrix leads to improved accuracy of the inferred planet parameters. 
When testing the use of scaling factors and offsets to try to correct for these systematic biases we found worse results than when relying on the covariance matrix. 
Thus we recommend that the covariance matrix always be published along with IFS data of exoplanets.

The systematic biases of the spectral extractions fundamentally limit the accuracy with which we can understand exoplanet atmospheres, with effects that can be much more significant than the statistical posterior precision.
One solution to the issues discussed in this work is to acquire higher quality data.  
Nevertheless, it will remain important to measure the covariance and to understand systematic effects imparted by data processing.

\begin{acknowledgements}
We would like to thank the anonymous referee for their insightful and detailed report, which substantially improved the quality of this work.
Software used: {\tt petitRADTRANS, pyKLIP, PynPoint, VIP-HCI, species,  pyMultiNest, Python, numpy, astropy, phot\_utils, matplotlib, scicomap.}
\end{acknowledgements}

%
%
\bibliographystyle{aa}
\bibliography{hr8799_gravity.bib}

\begin{appendix}
\section{SPHERE Data Reduction}\label{sec:sphereappendix}
We begin the process of extracting planetary spectra by applying a range of pre-processing steps, hereafter referred to as the data reduction stage.
Dark frames, detector flats and IFS flats are subtracted from the data. 
The spectral positions of each slice and the wavelengths are calibrated.
Bad pixels and cross talk are corrected, and the science frames are background subtracted.
ND filter transmission profiles are applied to stellar flux observations taken at the beginning and end of the ADI sequence observations.
To center the science frames, the satellite spots are used where available.
If available, satellite spots are used to calibrate the wavelength center of each channel.
These frames have anamorphism corrected and are shifted to a common center, and are output as a set of files with dimensions of $\left(x,y,\lambda\right)$.

For SPHERE data, we compare the results of using the standard SPHERE Data Center pipeline for the reduction to that of \cite{vigan-sphere-pipeline-2020}.
Recent updates to the pipeline have shown discrepancies in the wavelength solution for YJH data, however as this is only applicable to data taken in the satellite spot mode.
Lacking satellite spots, we instead rely on the standard ESO wavelength solution, and remain unaffected by the changes.

As it is challenging to inject fake companions into the raw detector frames for an IFS, we instead use the extracted spectrum of HR~8799~e for our comparison.
The raw frames were processed using both pipelines, and post-processed using \verb|KLIP| in order to extract the planetary spectrum.

Figure \ref{fig:SPHERE Data Reduction Comparison} shows the results of this extraction.
While qualitatively similar, the wavelength solutions are different between the pipelines, and both the location and amplitude of the 1.3 $\upmu$m feature disagree. 
Finally, there is significant discrepancy between the shapes of the spectra in the H band.
Without a clear metric for selecting a reduction pipeline, we choose to use the most up-to-date implementation of \cite{vigan-sphere-pipeline-2020} due to its ease of use and Python-based interface.

\begin{figure}[b]
    \centering
    \includegraphics[width=\linewidth]{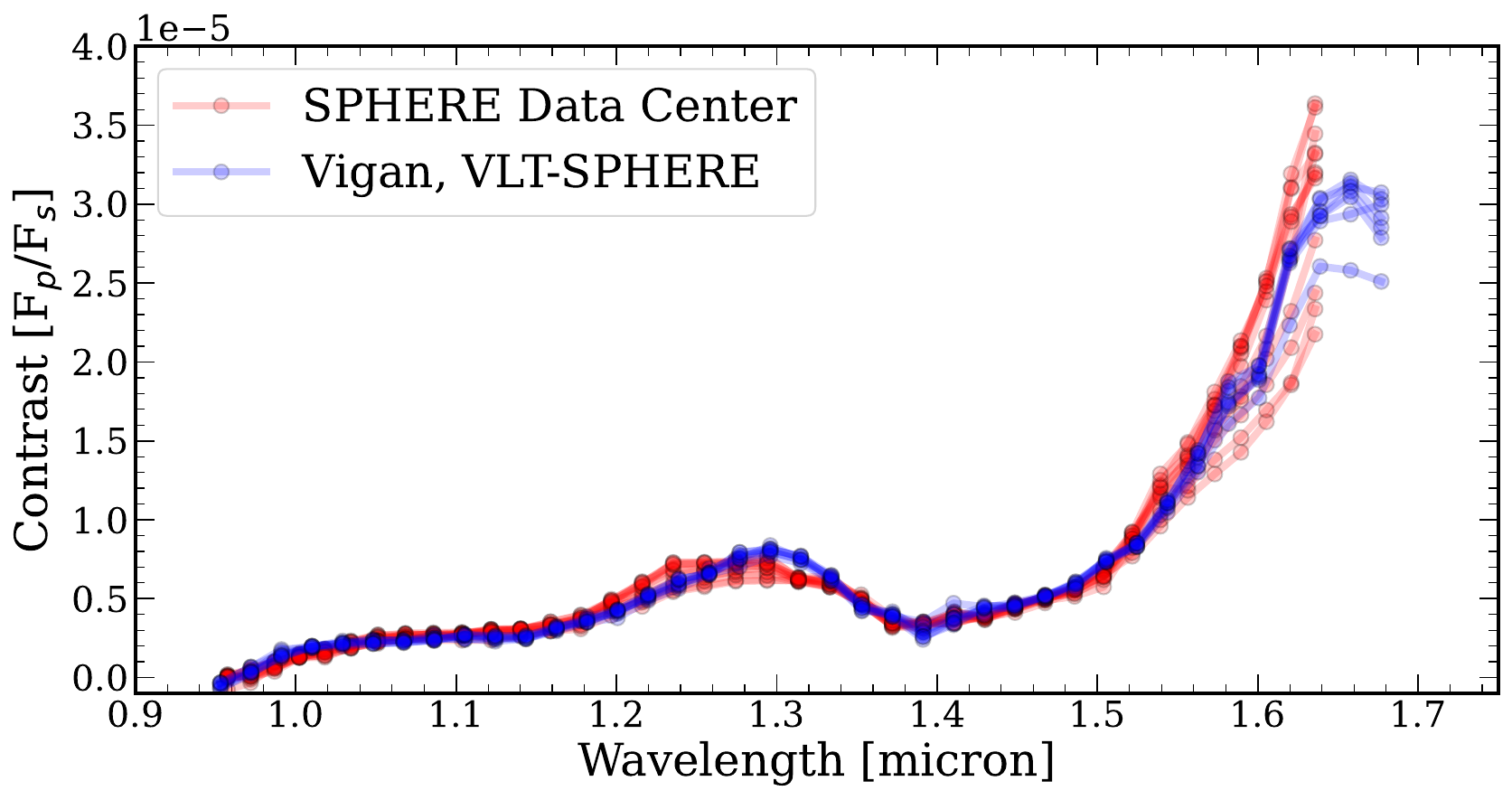}
    \caption{For the SPHERE data we compare the standard SPHERE Data Center data reduction (red) to the VLT-SPHERE pipeline described in \cite{vigan-sphere-pipeline-2020} (blue). 
    As planets cannot be injected into the raw data, we compare the spectrum of HR~8799~e as extracted with \texttt{KLIP}, where each measurement in the figure represents a different number of principal components used in the spectra extraction.}
    \label{fig:SPHERE Data Reduction Comparison}
\end{figure}
\end{appendix}
\end{document}